\documentclass[fleqn,usenatbib]{mnras}

\usepackage{newtxtext,newtxmath}

\usepackage[T1]{fontenc}

\DeclareRobustCommand{\VAN}[3]{#2}
\let\VANthebibliography\thebibliography
\def\thebibliography{\DeclareRobustCommand{\VAN}[3]{##3}\VANthebibliography}

\usepackage{graphicx}	
\usepackage{amsmath}	

\newcommand{\Zsun}{${\rm Z}_{\odot}$}
\newcommand{\cii}{{\sc [CII]~}}

\newcommand{\oiiio}{{\sc [OIII]} $\lambda \lambda 4959,5007$~}

\newcommand{\irxb}{${\rm IRX}$--$\beta$}
\newcommand{\irx}{${\rm IRX}$}
\newcommand{\muv}{${M_{\rm UV}}$}

\newcommand{\Msun}{${\rm M}_{\odot}$}

\newcommand{\mstar}{$M_{\star}$}

\title[REBELS-IFU: Dust attenuation curves]{REBELS-IFU: Dust attenuation curves of 12 massive galaxies at $\mathbf{z \simeq7}$}

\author[R. Fisher et al.]{R. Fisher,$^{1}$\thanks{E-mail: rebecca.fisher-7@postgrad.manchester.ac.uk (RF)}
R. A. A. Bowler,$^{1}$ 
M. Stefanon,$^{2, 3}$
L. E. Rowland,$^{4}$ 
H. S. B. Algera,$^{5, 6, 7}$ 
M. Aravena,$^{8}$ \newauthor
R. Bouwens,$^{4}$ 
P. Dayal,$^{9}$ 
A. Ferrara,$^{10}$
Y. Fudamoto,$^{11, 12}$
J. A. Hodge,$^{4}$
H. Inami,$^{6}$ 
K. Ormerod,$^{13}$\newauthor
A. Pallottini,$^{10}$
S. G. Phillips,$^{13}$
N. S. Sartorio,$^{14}$
R. Smit,$^{13}$
L. Sommovigo,$^{15}$
D. P. Stark,$^{12}$\newauthor
P. P. van der Werf$^{4}$
\\
$^{1}$Jodrell Bank Centre for Astrophysics, University of Manchester, Oxford Road, Manchester M13 9PL, UK\\
$^{2}$Departament d’Astronomia i Astrofìsica, Universitat de València, C. Dr. Moliner 50, E-46100 Burjassot, València, Spain \\
$^{3}$Unidad Asociada CSIC ”Grupo de Astrofísica Extragaláctica y Cosmología” (Instituto de Física de Cantabria - Universitat de València), Spain \\
$^{4}$Leiden Observatory, Leiden University, P.O. Box 9513, 2300 RA Leiden, The Netherlands \\
$^{5}$Institute of Astronomy and Astrophysics, Academia Sinica, 11F of Astronomy-Mathematics Building, No.1, Sec. 4, Roosevelt Rd, Taipei 106216, Taiwan, R.O.C. \\
$^{6}$Hiroshima Astrophysical Science Center, Hiroshima University, 1-3-1 Kagamiyama, Higashi-Hiroshima, Hiroshima 739-8526, Japan \\
$^{7}$National Astronomical Observatory of Japan, 2-21-1, Osawa, Mitaka, Tokyo, Japan \\
$^{8}$Instituto de Estudios Astrof\'{\i}sicos, Facultad de Ingenier\'{\i}a y Ciencias, Universidad Diego Portales, Av. Ej\'ercito 441, Santiago, Chile\\
$^{9}$Kapteyn Astronomical Institute, University of Groningen, P.O. Box 800, 9700 AV Groningen, The Netherlands \\
$^{10}$Scuola Normale Superiore, Piazza dei Cavalieri 7, 56126 Pisa, Italy \\
$^{11}$Center for Frontier Science, Chiba University, 1-33 Yayoi-cho, Inage-ku, Chiba 263-8522, Japan \\
$^{12}$Steward Observatory, University of Arizona, 933 N Cherry Avenue, Tucson, AZ 85721, USA \\
$^{13}$Astrophysics Research Institute, Liverpool John Moores University, 146 Brownlow Hill, Liverpool L3 5RF, UK\\
$^{14}$Sterrenkundig Observatorium, Krijgslaan 281 / S9, 9000 Gent, Belgium \\
$^{15}$Center for Computational Astrophysics, Flatiron Institute, 162 5th Avenue,
New York, NY 10010, USA \\
}

\date{Accepted XXX. Received YYY; in original form ZZZ}

\pubyear{2015}

\begin{document}
\label{firstpage}
\pagerange{\pageref{firstpage}--\pageref{lastpage}}
\maketitle

\begin{abstract}
We present measurements of the dust attenuation curves of 12 massive ($9~<~\log$(\mstar/\Msun)~$<~10$) Lyman-break galaxies at $z=6.5-7.7$ derived from \textit{James Webb Space Telescope (JWST)} NIRSpec integral field unit (IFU) spectroscopy.
The galaxies are drawn from the Atacama Large Millimeter/submillimeter Array (ALMA) Reionization Era Bright Emission Line Survey (REBELS) large program.
The dust attenuation curves were obtained by fitting spectral energy distribution (SED) models with a flexible dust law to the full galaxy spectra over observed wavelengths $0.6-5.3$~$\mu$m.
These attenuation curves show a range of recovered slopes ($-0.39\leq\delta\leq0.08$) that are on average slightly flatter than seen in local sources of the same stellar masses, with none exhibiting very steep slopes.
Three galaxies exhibit evidence for a 2175~{\AA} dust bump ($>4\sigma$) and we find SED fitting excluding the bump can overestimate derived stellar masses by up to $0.4$~dex. 
Correcting for the dust attenuation with our best-fit attenuation curves we recover a range of intrinsic UV-slopes ($-2.5\leq\beta_0\leq-2.2$).
The galaxies show moderate reddening ($A_V~=~0.1-0.6$~mag) and the $A_V$ to stellar mass relation is consistent with local sources.
The attenuation law slope is found to correlate with $A_V$, while we see no strong correlation with stellar mass, {\muv}, or gas-phase metallicity.
Overall, our results show little evolution in dust properties in the REBELS sources compared to the local Universe. 
Comparing our recovered trends to empirical models suggests that the most important factor driving the variation in the attenuation curves in our sample is the dust-star geometry, not the properties of the dust grains themselves.   
\end{abstract}




\section{Introduction}
In the last decade, the importance of dust in high-redshift galaxies ($z\gtrsim4$) has been revealed through the direct detection of the redshifted greybody emission from the dust grains themselves \citep[e.g.][]{Watson2015, Marrone2018, Hashimoto2019, Tamura2019, Harikane2020, Bakx2021, Akins2022,  Witstok2022, Mitsuhashi2023a, Algera2023, Algera2024}.
Atacama Large Millimeter/submillimeter Array (ALMA) programs such as the ALMA Large Program to INvestigate [CII] at Early times (ALPINE) targeting the dust content of normal star-forming galaxies at $4<z<6$ \citep[e.g.][]{ Bethermin2020, Faisst2020, LeFevre2020, Fudamoto2020} and the ALMA Reionization Era Bright Emission Line Survey (REBELS) large program at $z=6.5-8$ that targeted UV-bright galaxies \citep[{\muv}$<-21$,][]{Bouwens2022} have revealed that nearly half of the star formation in massive ($\log$(\mstar/\Msun)$ > 9$) galaxies at these redshifts is dust-obscured \citep[e.g.][]{Barrufet2023, Bowler2018, Bowler2023, Fudamoto2021, Inami2022, Schouws2022,  Algera2023}.
However, whether the dust properties at high redshift differ significantly from those in the local Universe is still unknown. 

Until recently, observations of $z\simeq7$ galaxies have almost exclusively used photometry that probes their rest-frame UV spectral energy distributions (SEDs).  
The colours of galaxies can be used to infer the presence of dust since dust grains absorb and scatter the UV and optical light produced by young stars causing reddening and attenuation \citep[e.g.][]{Meurer1999, Draine2003}. 
The rest-frame UV continuum is commonly parameterised using a power-law relation, $f_{\lambda} \propto \lambda^{\beta}$ \citep[e.g.][]{Calzetti1994, Meurer1999}.
A redder rest-frame UV-slope can thus imply a larger dust content. However, $\beta$ is also affected by the age, metallicity, and star formation history (SFH) of the stellar population \citep[e.g.][]{Bouwens2009b, Castellano2014} as well as the nebular continuum \citep[e.g.][]{Cullen2017, Reddy2018}.  Due to young stellar ages at early times, dust is suggested to be the dominant factor affecting the measured $\beta$-slope \citep{Wilkins2011, Tacchella2022}.  
Many studies have shown that the average $\beta$ value of galaxies decreases with increasing redshift, implying lower dust extinction \citep[e.g.][]{Bouwens2012, Dunlop2013, Topping2022b, Cullen2023, Austin2024, Topping2024b}, although some studies see no trend with redshift \citep[e.g.][]{Morales2023}.  
Recent work by \cite{Roberts-Borsani2024} using NIRSpec spectra supports the trend of bluer slopes at higher redshifts. 
The most massive galaxies at each redshift in these samples were found to have redder slopes suggesting these objects have a higher dust content.  
This is quantified by the colour-magnitude relation, $\beta-M_{UV}$, with redder UV-slopes being found in the higher luminosity, more massive star-forming galaxies, suggesting they have higher dust attenuation and/or metallicities \citep{Reddy2010, Bouwens2014, Yamanaka2019, Mitsuhashi2023a, Bowler2023}.

At lower redshifts the dust attenuation curve describing the wavelength-dependent effect of dust absorption and scattering of stellar photons has been inferred.
The shape of this curve depends on the chemical composition of the dust, the distribution of dust grain sizes, and the dust-star spatial geometry (see \citeauthor{Salim2020} \citeyear{Salim2020} for a review).
The attenuation curve can be obtained using spectroscopic observations of the Balmer emission lines and comparing the average SEDs of galaxies with different Balmer decrements \citep[e.g.][]{Calzetti1994, Reddy2015, Battisti2016, Battisti2017, Shivaei2020, Battisti2022, Cooper2024}. 
Alternatively, the attenuation curve can be obtained by comparing the SED fitted to observations to an assumed intrinsic spectrum \citep{Cullen2018} or SED fitting with a flexible dust attenuation law \citep[e.g.][]{Salim2018, Boquien2022, Markov2023, Markov2024}.

Studies such as \cite{Battisti2022} suggest that the attenuation curve evolves continuously with redshift, since they find that the average dust attenuation curve slope and normalisation at $z\simeq1.3$ is between the \cite{Calzetti2000} curve from local starburst galaxies and the steeper curve of \cite{Reddy2015} at $z\simeq2$. 
However, the attenuation curve slope has also been found to depend on properties other than redshift, complicating the interpretation. 
For example, \cite{Shivaei2020} find that the slope of the dust attenuation curve at $z\simeq2$ may be metallicity dependent, with lower metallicity galaxies exhibiting steeper slopes more similar to that of the Small Magellanic Cloud (SMC) extinction curve.  
In contrast, at $z<0.3$, \cite{Salim2018} find that the attenuation curves for individual galaxies have a wide variety of slopes with some dependence on stellar mass but no clear trend with metallicity.  
Additionally, more massive galaxies with older stellar populations, higher metallicities, higher star formation rates (SFRs), and higher dust attenuation magnitudes have been seen to show shallower, more Calzetti-like, dust attenuation curves \citep{Cullen2018, McLure2018}, while lower mass galaxies may have steeper, more SMC-like curves \citep{Reddy2010, Bouwens2016, Reddy2018, Shivaei2020, Shivaei2020b}.

At $z\simeq5$, \cite{Boquien2022} also find a range of attenuation curve slopes ($-1.84 \leq \delta \leq0.23$) in a subset of the ALPINE galaxies, demonstrating that a single attenuation curve is not appropriate for these main-sequence galaxies.  
The shape of the attenuation curve may also deviate significantly from what is seen locally, as inferred from NIRSpec observations of HI recombination lines for a galaxy at $z=4.41$ by \cite{Sanders2024b}. 
These findings suggest that dust grain properties and/or dust-star geometry could differ at high redshift compared to local star-forming galaxies and can also significantly vary between galaxies.
This has implications for derived properties, for example, \cite{Reddy2015} find that the assumption of the wrong attenuation curve can cause SFR and mass differences of up to $20$\% and 0.16 dex, respectively. 

With the \textit{James Webb Space Telescope (JWST)} we now have the spectroscopic sensitivity and wavelength coverage to measure both the shape and magnitude of dust attenuation for the first time at very high redshifts. 
For example, recent work by \cite{Markov2024} measuring attenuation curves at $z=2-12$ with NIRSpec found that attenuation curve shapes may significantly deviate from those in the local universe, causing deviations in SFR and mass estimates of up to 0.4 dex compared to the typically assumed \cite{Calzetti2000} law.
This study also finds tentative evidence that the slopes flatten with redshift and attributes this to larger dust grains at earlier epochs since there has been insufficient time for reprocessing in the interstellar medium (ISM) to produce smaller grains.  
Detailed measurements of individual galaxies have revealed evidence for the 2175~{\AA} dust bump in the Epoch of Reionisation \citep{Witstok2023, Markov2023, Markov2024}.
This bump feature is believed to be associated with small carbonaceous grains \citep[polycyclic aromatic hydrocarbons (PAHs); see][]{Draine2003}, implying that these grains are rapidly produced and survive in the ISM of high redshift galaxies \citep[see][for a review]{Schneider2023}.

In this work, we measure the dust attenuation curves directly from a sample of massive Lyman-break galaxies at $z\simeq7$.
This subset of the REBELS galaxies is a unique dataset for which we have NIRSpec spectra at rest-UV and rest-optical wavelengths with a spectral resolution $R\simeq100$.  
This allows us to measure the gas-phase metallicity from optical emission lines (Rowland et al. in prep.) and estimate more robust stellar masses (Stefanon et al. in prep.).
The galaxies at these redshifts in studies to date have lower stellar masses \citep[e.g.][]{Witstok2023, Markov2023, Markov2024}, whereas the REBELS sample probes the massive end with larger inferred dust contents and, thus, dust attenuation is expected to have a more significant effect on observed properties.
The REBELS sample also uniquely benefits from FIR spectral coverage from ALMA observations that can be combined with the analysis of the \textit{JWST} observations presented here. 
Constraints on dust production models can also now be made using the dust-to-gas, dust-to-metal, and dust-to-stellar ratios for these galaxies (Algera et al. in prep.).

The structure of this paper is as follows. In Section \ref{sec:data} we introduce the REBELS IFU sample and dataset.  In Section \ref{sec:methods} we describe the SED fitting setup with a flexible dust attenuation law and how physical properties were extracted from the galaxy spectra.  In Section \ref{sec:results} we present the results from these fits including the dust attenuation curves for each galaxy.
We discuss the implications of our results on the dust properties of these sources in Section \ref{sec:discussion}.
The key findings are summarised in Section \ref{sec:summary}.  
We assume the standard $\Lambda$CDM cosmology with $H_0 = 70$ km s$^{-1}$ Mpc$^{-1}$, $\Omega_{\text{m}} = 0.3$, and $\Omega_{\Lambda} = 0.7$ throughout this work \citep{Planck2020}.

\begin{table*}
	\centering
	\caption{The 12 REBELS galaxies targeted with the NIRSpec IFU observations used in this study.  
    Col.\ (1): Galaxy identifier.
    Col.\ (2): Spectroscopic redshift from the {\cii}detection from ALMA \citep{Bouwens2022}. 
    Col.\ (3): Rest-frame UV-continuum slope, $\beta$.
    Col.\ (4): Absolute rest-frame UV magnitude at $1500$ {\AA}, {\muv}.
    Col.\ (5): Metallicities, $Z$, derived from optical emission lines (Rowland et al. in prep.).
    The best-fit parameters from the $\tt{BAGPIPES}$ SED fits with a flexible dust attenuation law are then shown.
    Col.\ (6): V-band continuum attenuation, $A_V$.
    Col.\ (7): Slope of the dust attenuation curve expressed as the deviation, $\delta$, of the power-law exponent from the Calzetti-like curve. 
    Col.\ (8): The 2175~{\AA} dust bump strength, $B$.
    Col.\ (9): Stellar mass, {\mstar}. Note that these differ from those presented in Stefanon et al. (in prep.) who assume a fixed Calzetti dust attenuation curve, but are consistent within the errors.
    Col.\ (10): The intrinsic UV-slope, $\beta_0$, of the SED model (see Section~\ref{sec:intrinsic_spectra}).}
	\label{tab:pt1}
	\begin{tabular}[]{ccccccccccc} 
        \hline
        ID & $z$ & $\beta$ & {\muv} & $Z$ & $A_V$ & $\delta$ & $B$ & $\log_{10}$(\mstar/\Msun) & $\beta_0$ \\
        &  &  & /mag & /{\Zsun} & /mag &  &  &  &  \\
        (1) & (2) & (3) & (4) & (5) &  (6) & (7) & (8) & (9) & (10) \\
        \hline
        REBELS-05 & 6.496 & $-1.42 \pm 0.05$ & $-21.49 \pm 0.07$ & $0.66 \pm 0.25$ & $0.38\substack{+0.10 \\ -0.09}$ & $-0.18\substack{+0.06 \\ -0.07}$ & $0.18\substack{+0.17 \\ -0.12}$ & $9.60\substack{+0.11 \\ -0.10}$ & $-2.31 \pm 0.06$  \\
        REBELS-08 & 6.749 & $-1.92 \pm 0.05$ & $-21.88 \pm 0.03$ & $0.34 \pm 0.16$ & $0.28\substack{+0.07 \\ -0.07}$ & $-0.10\substack{+0.06 \\ -0.08}$ & $3.00\substack{+0.48 \\ -0.40}$ & $9.30\substack{+0.11 \\ -0.10}$ & $-2.41 \pm 0.05$  \\
        REBELS-12 & 7.346 & $-1.68 \pm 0.03$ & $-22.39 \pm 0.03$ & $0.35 \pm 0.11$ & $0.18\substack{+0.08 \\ -0.09}$ & $-0.31\substack{+0.20 \\ -0.29}$ & $0.15\substack{+0.22 \\ -0.10}$ & $9.80\substack{+0.09 \\ -0.10}$ & $-2.37 \pm 0.04$  \\
        REBELS-14 & 7.084 & $-1.74 \pm 0.03$ & $-22.30 \pm 0.04$ & $0.16 \pm 0.05$ & $0.17\substack{+0.07 \\ -0.06}$ & $-0.39\substack{+0.14 \\ -0.23}$ & $0.24\substack{+0.27 \\ -0.15}$ & $9.54\substack{+0.12 \\ -0.14}$ & $-2.44 \pm 0.05$  \\
        REBELS-15 & 6.875 & $-2.01 \pm 0.03$ & $-22.40 \pm 0.03$ & $0.12 \pm 0.09$ & $0.35\substack{+0.08 \\ -0.07}$ & $-0.03\substack{+0.04 \\ -0.04}$ & $0.99\substack{+0.15 \\ -0.14}$ & $9.40\substack{+0.03 \\ -0.03}$ & $-2.45 \pm 0.03$  \\
        REBELS-18 & 7.675 & $-1.56 \pm 0.03$ & $-22.11 \pm 0.02$ & $0.64 \pm 0.19$ & $0.27\substack{+0.04 \\ -0.05}$ & $-0.39\substack{+0.07 \\ -0.12}$ & $0.56\substack{+0.18 \\ -0.16}$ & $9.98\substack{+0.04 \\ -0.04}$ & $-2.39 \pm 0.03$  \\
        REBELS-25 & 7.307 & $-1.61 \pm 0.09$ & $-21.46 \pm 0.05$ & $0.85 \pm 0.33$ & $0.25\substack{+0.09 \\ -0.06}$ & $-0.35\substack{+0.13 \\ -0.15}$ & $2.72\substack{+0.65 \\ -0.67}$ & $9.07\substack{+0.10 \\ -0.08}$ & $-2.22 \pm 0.12$  \\
        REBELS-29 & 6.685 & $-1.89 \pm 0.05$ & $-22.00 \pm 0.04$ & $1.11 \pm 0.40$ & $0.33\substack{+0.08 \\ -0.07}$ & $-0.09\substack{+0.07 \\ -0.08}$ & $0.71\substack{+0.30 \\ -0.27}$ & $9.94\substack{+0.06 \\ -0.08}$ & $-2.38 \pm 0.06$  \\
        REBELS-32 & 6.729 & $-1.34 \pm 0.07$ & $-21.16 \pm 0.08$ & $0.61 \pm 0.19$ & $0.46\substack{+0.14 \\ -0.13}$ & $-0.08\substack{+0.07 \\ -0.09}$ & $0.53\substack{+0.29 \\ -0.29}$ & $9.75\substack{+0.11 \\ -0.12}$ & $-2.38 \pm 0.07$  \\
        REBELS-34 & 6.634 & $-2.23 \pm 0.03$ & $-22.25 \pm 0.02$ & $0.44 \pm 0.32$ & $0.10\substack{+0.05 \\ -0.04}$ & $-0.14\substack{+0.13 \\ -0.19}$ & $1.03\substack{+0.52 \\ -0.51}$ & $9.59\substack{+0.09 \\ -0.07}$ & $-2.31 \pm 0.04$  \\
        REBELS-38 & 6.577 & $-1.63 \pm 0.06$ & $-21.99 \pm 0.05$ & $0.39 \pm 0.17$ & $0.58\substack{+0.11 \\ -0.14}$ & $0.08\substack{+0.05 \\ -0.05}$ & $0.12\substack{+0.13 \\ -0.08}$ & $9.93\substack{+0.08 \\ -0.09}$ & $-2.47 \pm 0.06$  \\
        REBELS-39 & 6.845 & $-2.07 \pm 0.04$ & $-22.39 \pm 0.04$ & $0.21 \pm 0.16$ & $0.16\substack{+0.06 \\ -0.04}$ & $-0.23\substack{+0.08 \\ -0.10}$ & $0.94\substack{+0.39 \\ -0.35}$ & $9.56\substack{+0.11 \\ -0.09}$ & $-2.37 \pm 0.04$  \\
        \hline
	\end{tabular}
\end{table*}

\section{Data and sample}
\label{sec:data}
The galaxies studied in this work are a subset of the 40 massive \citep[$\log_{10}(M_*/M_\odot) = 8.8-10.4$,][]{Topping2022} \footnote{Note that for the galaxies studied in this work the stellar masses are re-derived from the SED fits to the \textit{JWST} NIRSpec spectra.} Lyman-break galaxies at $z=6.5-9.5$ targeted by the REBELS ALMA large program.  This survey aimed to detect dust continuum and {\cii}$158 \mu$m emission in bright, star-forming galaxies identified from wide-field ground-based imaging and \textit{Hubble Space Telescope (HST)} archival fields.  A detailed summary of the selection and results from the first year of observations of the REBELS galaxies can be found in \cite{Bouwens2022}.

\textit{JWST} NIRSpec observations were obtained (PID 1626; P.I. Stefanon and PID 2659; P.I. Weaver, \textit{JWST} Cycle 1) for the 12 galaxies listed in Table~\ref{tab:pt1}. 
These galaxies were selected based on having the brightest {\cii}$158 \mu$m emission in the REBELS program and SFRs inferred from {\cii}of $50-400$~{\Msun}/yr \citep{DeLooze2014}.
These SFRs are significantly higher than typical galaxies at these redshifts (although we note that the SFRs inferred from UV+IR estimates tend to be slightly lower, e.g. \citeauthor{Bowler2023} \citeyear{Bowler2023}).
They are a representative sample of the ranges of physical properties, such as stellar mass, rest-UV slope, and {\sc [OIII]}$+$H$\beta$ equivalent width, seen in the full REBELS sample.

The IFU observations consist of a spectrum between wavelengths of $0.6-5.3$~$\mu$m for each pixel in the $3.1''\times3.2''$ field of view.
We used the prism mode with a resolution of $R\simeq100$ to efficiently measure both the rest-UV/optical continuum shapes and bright line emission.
An exposure time of approximately 30 minutes per source was chosen to provide high S/N spatially-resolved information in the rest-frame optical (S/N $\sim8-10$) and for key optical emission lines (S/N $ >10$).  
To obtain a total galaxy spectrum, a source aperture for each galaxy was created by combining the 5$\sigma$ isophotal contours in wavelength ranges corresponding to H$\alpha$, H$\beta$, UV continuum, optical continuum, {\sc [OII]} $\lambda \lambda 3727,3729$, and {\oiiio}emission.
This approach accounts for any wavelength dependence in the morphology.
There was good consistency between the spectrum extracted from these pixels and previous photometry \citep{Bouwens2022}. 
We find that the default error extension on the cube underestimates the error and, thus, we use the root mean square (RMS) of the extracted spectra instead. 
The full details of the cube data reduction will be presented in a future paper (Stefanon et al. in prep.).

\section{Methods}
\label{sec:methods} 

\begin{figure} 
\includegraphics[width=\columnwidth]{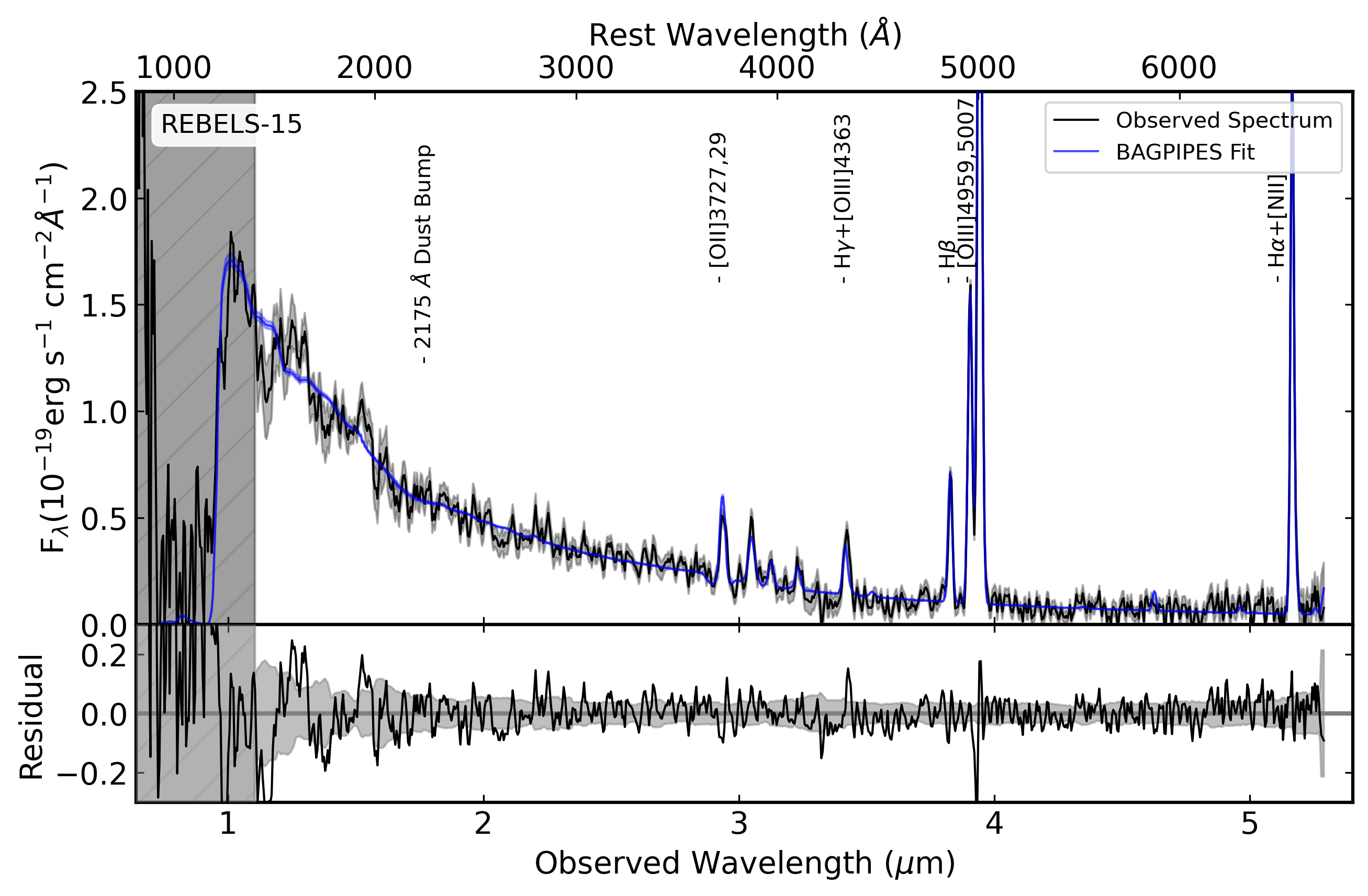}
\caption{An example of the full NIRSpec PRISM spectrum for one of the REBELS IFU galaxies with the 1$\sigma$ error shown by the grey shaded region.  The best-fitting SED model using a flexible dust attenuation law is shown in blue.  The spectrum is masked below a rest-frame wavelength of 1400~{\AA} (grey hatched region) to exclude effects caused by potential Ly$\alpha$ damping and instrumental effects at low wavelengths.  The bottom panel shows the fit residuals.  REBELS-15 shows evidence for a 2175~{\AA} dust bump at $\sim1.7 \mu$m.}
\label{example_spectrum_fit} 
\end{figure}

\begin{table}
	\centering
	\caption{The $\tt{BAGPIPES}$ SED fitting parameters and their priors used in this work.  These are the redshift ($z$), stellar mass ({\mstar}), metallicity ($Z$), ionisation parameter ($U$), V-band continuum attenuation ($A_V$), multiplicative factor on $A_V$ for the emission lines ($\eta$), the dust law slope ($\delta$), and the $2175$~{\AA} bump strength ($B$). }
	\label{tab:BAGPIPES_priors}
	\begin{tabular}[]{ccccccccccccr} 
        \hline
        Parmaeter & Limits & Prior\\
        \hline
        $z$ & -- & Fixed\\
        $\log_{10}(${\mstar}/{\Msun}) & (7, 12) & Uniform \\
        $Z$ / {\Zsun} & (1e-06, 10) & Logarithmic \\
        $\log U$ & (-3, 0) & Uniform \\
        $A_V$ / mag & (0, 5) & Uniform \\
        $\eta$ & (1, 3) & Uniform \\
        $\delta$ & (-2, 0.75) & Uniform \\
        $B$ & (0, 4) & Uniform \\
        \hline
	\end{tabular}
\end{table}

We infer the best-fitting dust attenuation laws through SED fitting. The fits were performed using $\tt{BAGPIPES}$ \citep{Carnall2018, Carnall2019}.
An example fit is shown in Fig.~\ref{example_spectrum_fit} and the best-fit parameters for each galaxy are shown in Table~\ref{tab:pt1}. 
A summary of the SED fitting parameters and the priors we used can be found in Table~\ref{tab:BAGPIPES_priors}.
We used a non-parametric SFH model with a continuity prior.  
The SFH consists of six bins.  The two most recent are set to $3$~Myr and $7$~Myr.   The remaining four are evenly distributed in log space up to $z=20$.
Non-parametric models are more flexible and have been found to better reproduce the shape of more complex SFHs with fewer biases and smaller systematic uncertainties on the derived parameters than parametric models \citep{Leja2019, Lower2020, Topping2022, Markov2023, Whitler2023}. 
Our key results are unchanged if we assume a parametric SFH (see Appendix~\ref{sec:appendix}).
We use the \cite{Bruzual2003} stellar population model, which uses a \cite{Kroupa2001} initial mass function (IMF).
The grids we use for the nebular line and continuum emission, generated using $\tt{CLOUDY}$ \citep{Ferland2017}, allow the $\log U$ parameter to vary between (-3, 0). 
To check the derived dust laws are not affected by systematics originating from our choice of stellar model, we also ran using BPASS stellar population models \citep{Stanway2018} and found the dust attenuation curves are consistent within the errors, with no clear biases (see Appendix~\ref{sec:appendix}). 

From independent emission line fitting (Rowland et al. in prep), we find that the line dispersion of the NIRSpec spectra is lower than the pre-flight expectation.
Therefore, we use an updated line-spread function in the fitting to account for the variable spectral resolution of the NIRSpec PRISM.  
The redshifts were fixed to the {\cii}spectroscopic values.
Small redshift offsets ($\Delta z \leq 0.01$) from the {\cii}spectroscopic redshifts found during the emission line fits are accounted for by shifting the observed wavelength array of the spectra. 
Although direct measurements of the metallicities from emission lines using strong-line calibrations have been made (Rowland et al. in prep.), we choose to leave the metallicity as a free parameter with a logarithmic prior to minimize potential systematic effects in the SED fitting arising from the limited number of templates.  
Additionally, we mask the spectrum at wavelengths below 1400~{\AA} to exclude the Ly$\alpha$ damping region \citep{Heintz2024} and instrumental effects at short wavelengths.  

In our fits, we use the flexible dust attenuation law model in $\tt{BAGPIPES}$ that follows the parameterisation of \cite{Salim2018}. 
This model allows flexibility in the slope of the dust law but can also recover the commonly used curves (e.g. Calzetti-like and SMC).
The shape of the dust attenuation curve is parameterised by  
\begin{eqnarray}
    \frac{A_\lambda}{A_V} = \left(\frac{\lambda}{5500 \text{\AA}}\right)^{-n} + \frac{D_{\lambda}(B)}{R_{V}}
    \label{eq:Salim_dust},
\end{eqnarray}  
where $n = -\delta + n_{\rm Calzetti} = -\delta + 0.75$ such that $\delta$ is the deviation from the Calzetti-like attenuation curve (i.e. for a Calzetti-like curve $\delta=0$ and for the SMC extinction curve $\delta\simeq-0.45$), $A_{\lambda}$ is the attenuation magnitude at the wavelength $\lambda$, and $A_V$ is the $V$-band (5500~{\AA}) attenuation magnitude \citep{Salim2018}. 
The second term describes the dust bump centred at $\lambda_0=2175$~{\AA} that is modelled by a Drude profile
\begin{eqnarray}
    D_{\lambda}(B) = \frac{B \lambda^2 w^2}{(\lambda^2 - \lambda_0^2)^2 + \lambda^2w^2}
    \label{eq:drude_bump},
\end{eqnarray}
with width $w=350$~{\AA} and amplitude $B$ in units of $A_{\text{bump}}/E(B-V)$.
The uncertainties on the dust attenuation curves were calculated from the errors on $\delta$ and $B$ provided by $\tt{BAGPIPES}$.
The total-to-selective extinction ratio, $R_V$, is given by 
\begin{eqnarray}
    R_V = \frac{A_V}{E(B-V)} = \frac{A_V}{A_B-A_V}
    \label{eq:Rv_expression},
\end{eqnarray} 
where $A_B$ is the $B$-band (4400~{\AA}) attenuation.  
Using this we can obtain the total attenuation curve as
\begin{eqnarray}
    k(\lambda) = \frac{A_{\lambda}R_V}{A_V}
    \label{eq:k_lambda_from_AlambdaAv}.
\end{eqnarray} 
The multiplicative factor on $A_V$ for stars in birth clouds, $\eta$, was also allowed to vary since measurements of nebular attenuation from the Balmer emission lines imply this may differ from the continuum attenuation (Fisher et al. in prep.). 
In practice, this means that the dust attenuation applied to the emission lines in the SED fit equals $\eta$ multiplied by the continuum attenuation, $A_V$.
Allowing this parameter to vary does not significantly change the derived shape of the dust attenuation curves, however, it does result in lower continuum $A_V$ values than fits with $\eta=1$.  
To test the validity of this approach, we ran the SED fitting on simulated galaxy spectra and found the dust attenuation curve is accurately recovered, with smaller errors on the dust attenuation law slope for galaxies with higher $A_V$ values, as we would expect.  This is shown in Fig.~\ref{SED_fit_tests} and discussed in Appendix~\ref{sec:appendix}.

The rest-frame UV-slope, $\beta$, was obtained directly from the observed spectra by fitting a power-law between the rest-frame wavelengths $\lambda_{\text{rest}}= 1268-2580$~{\AA} using a least-squares fitting method.  The values are consistent within the errors with those obtained if we fit using only the Calzetti windows \citep{Calzetti1994}, implying the rest-UV region is not affected by strong emission or absorption lines in these sources.  
The monochromatic rest-frame UV magnitude, {\muv}, was calculated from the spectrum flux, $F_{\lambda}$, in a top hat filter of width 100 {\AA} centred at $\lambda_{\text{rest}}=1500$ {\AA}.

\begin{figure*}
\includegraphics[width=2\columnwidth]{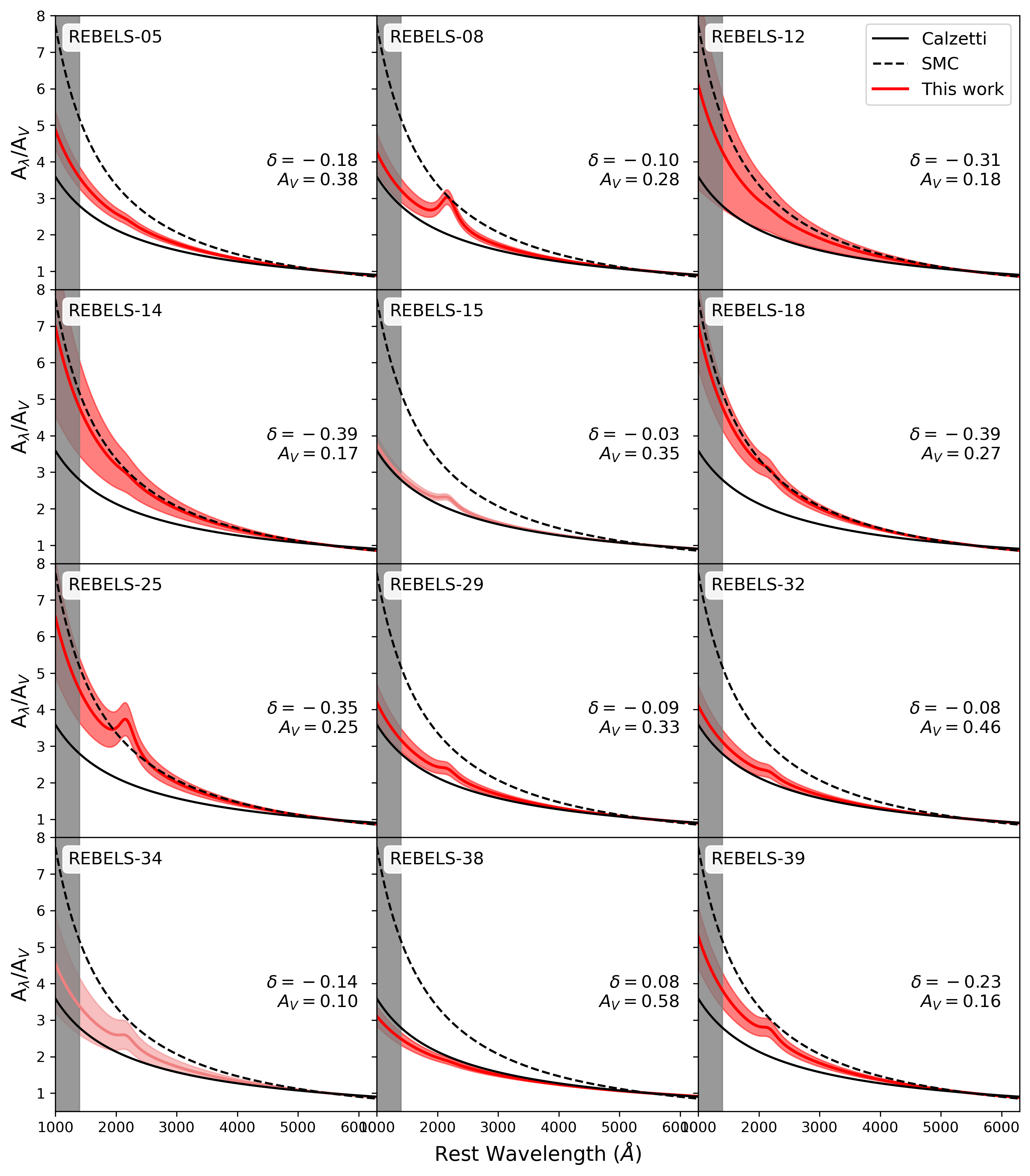}
\caption{The best-fit dust attenuation curves for 12 massive galaxies at $z\simeq7$ from the REBELS IFU program. These curves were obtained from SED fits to the global NIRSpec spectra extracted from the IFU.  We use $\tt{BAGPIPES}$ to fit the SED models with a non-parametric SFH and a \citet{Salim2018} flexible dust law parametrisation.  The observed spectrum was masked below a rest wavelength of 1400~{\AA}, as shown by the grey-shaded band. The red-shaded region shows the 1$\sigma$ error on the attenuation curves. The two galaxies with no dust continuum detections (REBELS-15 and REBELS-34) are shown in a fainter shade of red.  The Calzetti-like (SMC) curve is shown by the solid (dashed) black line.}
\label{BAGPIPES_dust_law_fits} 
\end{figure*}

\section{Results}
\label{sec:results} 
\subsection{Dust attenuation curves} 
The best-fitting dust attenuation curve for each galaxy is shown in red in Fig.~\ref{BAGPIPES_dust_law_fits}.
The curves exhibit a variety of slopes and, given the size of the shaded error regions, there are distinguishable differences between them and standard assumed curves like the SMC extinction relation. 
The $\delta$ values shown in Table~\ref{tab:pt1} are used to quantify the slopes of the curves with respect to the Calzetti-like curve ($\delta=0$).
REBELS-38 has a positive $\delta$ value and is the only curve shallower than the Calzetti-like curve.
Four attenuation curves (REBELS-15, REBELS-29, REBELS-32, and REBELS-34) are consistent within the errors with the Calzetti-like relation.
Steeper curves lying between the Calzetti-like and SMC relations are seen in three sources (REBELS-05, REBELS-08, and REBELS-39).
REBELS-12 has a steeper curve but the error is such that it could be consistent with either the Calzetti-like or SMC relation. 
The remaining three galaxies (REBELS-14, REBELS-18, and REBELS-25) have the steepest curves, consistent within the errors with the SMC relation with $\delta\simeq-0.45$.
The mean slope of the sample is $\delta=-0.18\pm0.15$, which is consistent with the majority of the curves lying closer to the Calzetti-like curve than the SMC, as expected for more massive, higher metallicity galaxies \citep[e.g.][]{Cullen2018, McLure2018, Shivaei2020}.
Indeed, most are inconsistent with being as steep as the SMC curve. 
The three attenuation curves with the largest errors on $\delta$ are REBELS-12, REBELS-14, and REBELS-25 which all reside at $z>7$.  This means more of the rest-optical region of the spectrum is redshifted beyond the wavelength coverage of NIRSpec.
The error on REBELS-34 is also larger than the other galaxies since it has the lowest $A_V$ value, meaning dust has less of an impact on the spectrum (see Appendix~\ref{sec:appendix}).
This is unsurprising given the non-detection of the FIR dust continuum with ALMA for REBELS-34 \citep{Inami2022}.

Three galaxies (REBELS-08, REBELS-15, and REBELS-25) have bump strengths greater than zero at more than $4\sigma$ significance ($7.5\sigma$, $7.2\sigma$, and $4.0\sigma$, respectively).
The dust attenuation curve of REBELS-08 shows evidence for a strong 2175~{\AA} bump, with a recovered bump strength of $B~=~3.00\substack{+0.48 \\ -0.40}$. 
For comparison, the bump strength in the Milky Way extinction curve has a mean value of $B=3$ \citep{Salim2018}. 
REBELS-25 also has a high recovered bump strength of $2.72\substack{+0.65 \\ -0.67}$, although this attenuation curve is less well constrained, likely due to its higher redshift reducing the wavelength coverage of the rest-optical spectrum and the lower S/N of the spectrum. 
If we fix the bump strength to $B=0$ in the fits for REBELS-08 and REBELS-25 we recover dust curves with shallower slopes.
We compare the Bayesian information criterion (BIC) values to help assess which model is preferred.
REBELS-08 and REBELS-25 are the only two galaxies for which the fits with a bump term exhibit significantly lower BIC values ($\Delta$BIC $ = 65$ and $15$, respectively), indicating the model is preferred over the fits with no bump term.
A more detailed analysis of dust bumps from the REBELS-IFU data will be presented in Ormerod et al. (in prep.).

To allow for more variety in dust attenuation law shape we also fit using the \cite{Li2008} parameterisation that has four free parameters.  
We find that the shapes agree well with the Salim model fits.
The BIC values were similar in all cases with a slight preference for the \cite{Salim2018} model ($\Delta$BIC $=0-16$).
Thus, there is no statistical evidence that the more complex model better describes the shapes of the attenuation curves, which might have implied that the shapes significantly deviate from local relations, and so we only present our results using the \cite{Salim2018} model.  
We caveat that the reduced wavelength coverage in the FUV of our spectra compared to those used in \cite{Markov2023, Markov2024} could be partially responsible for this.

\subsection{Dust law slope dependence on physical properties}
\label{sec:results_4.2}
\begin{figure*} 
\includegraphics[width=\columnwidth]{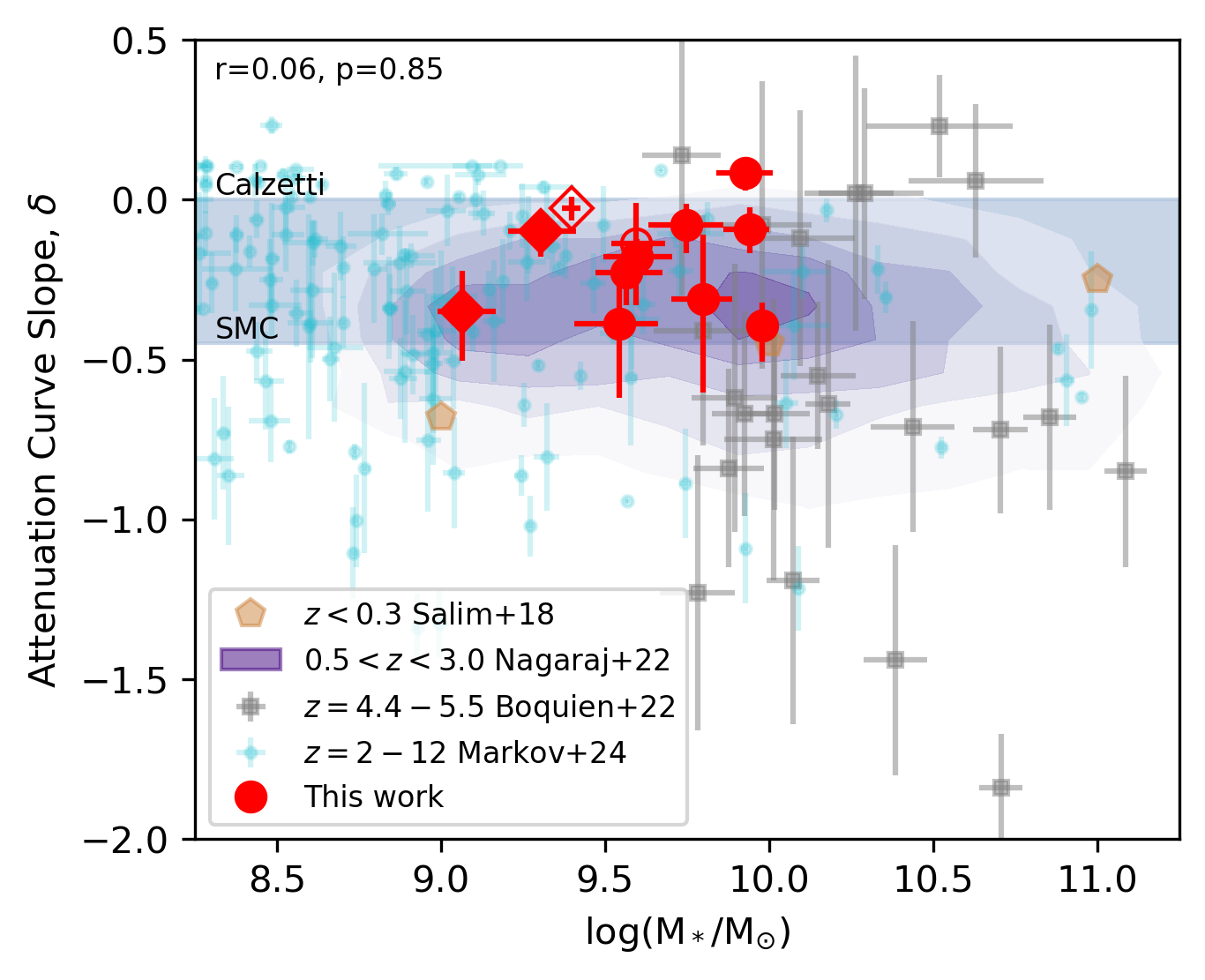}
\includegraphics[width=\columnwidth]{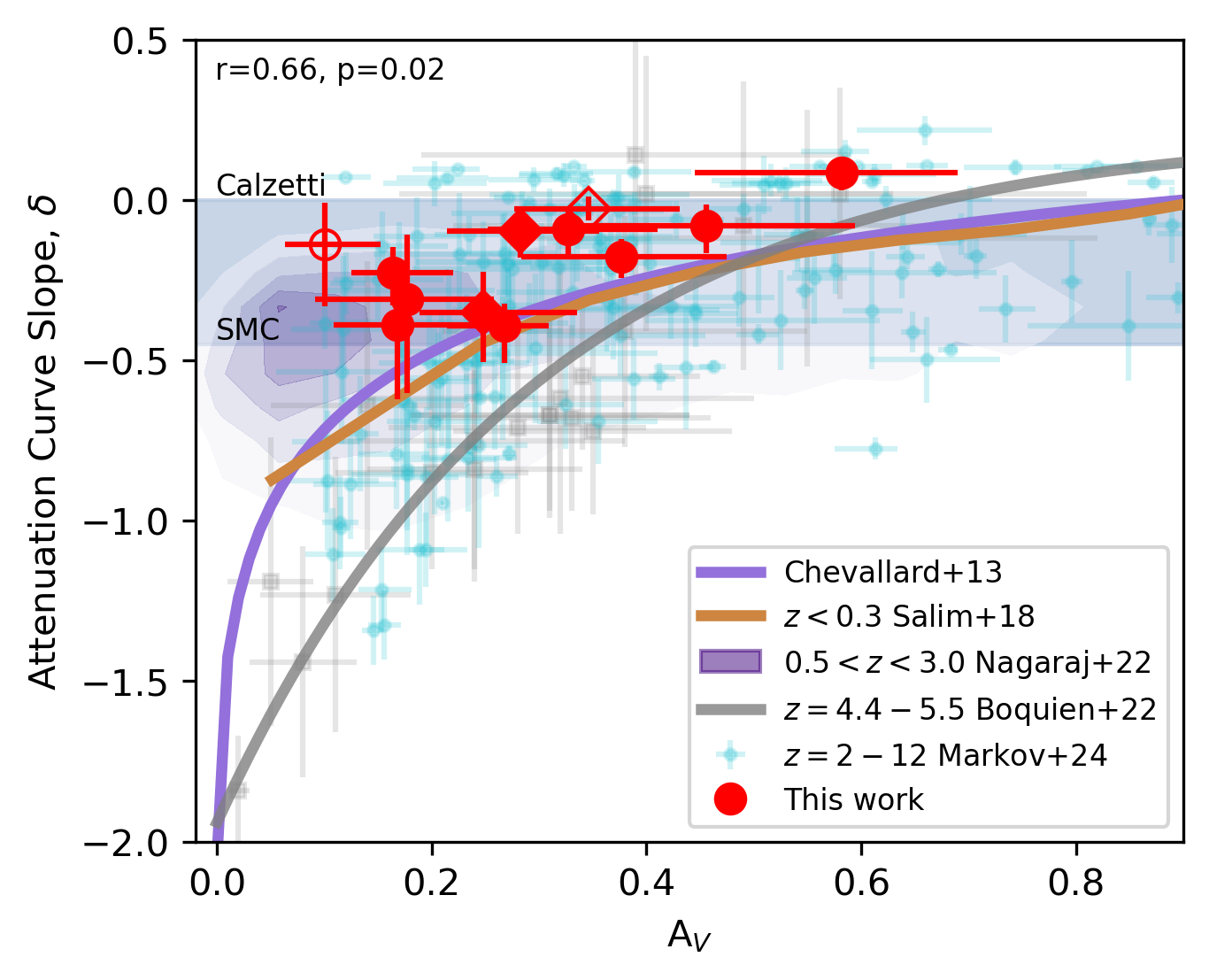}
\includegraphics[width=\columnwidth]{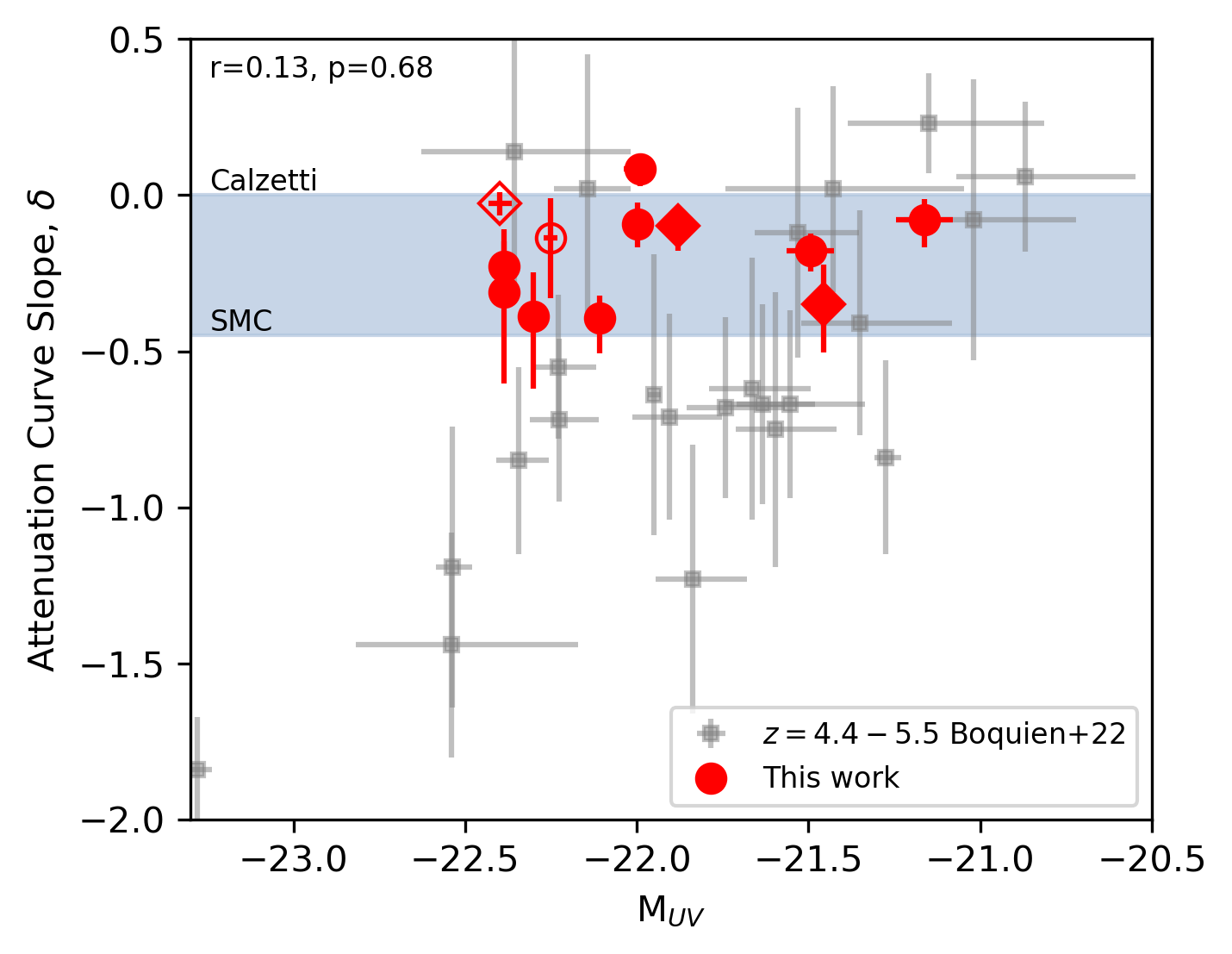}
\includegraphics[width=\columnwidth]{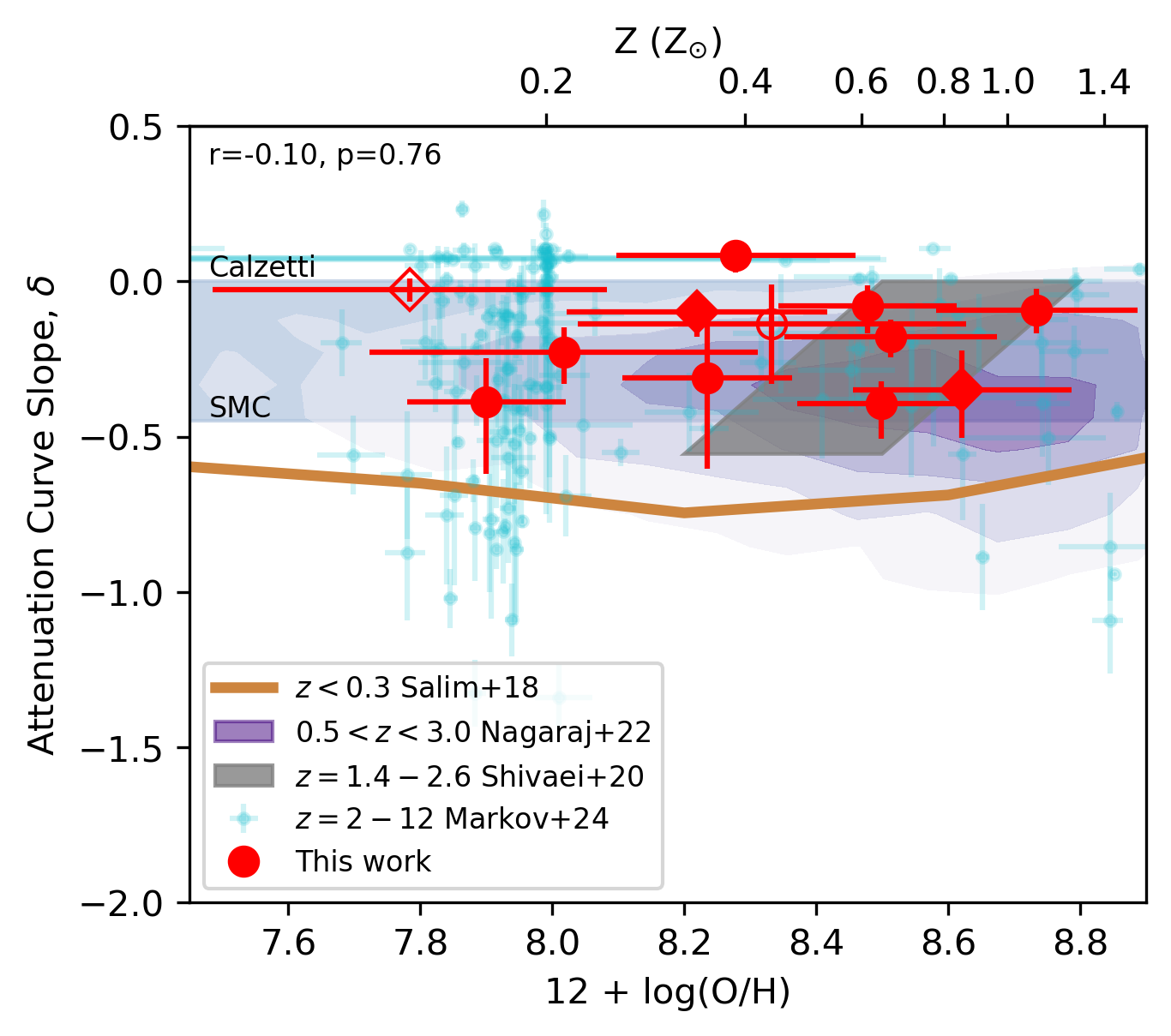}
\caption{The derived dust attenuation curve slopes of the 12 massive galaxies from the REBELS IFU sample at $z\simeq7$ plotted against physical properties.  The slope is expressed as the deviation, $\delta$, of the power-law exponent from the Calzetti-like curve.  The slope correlates with $A_V$ (top right), but we find no clear correlations with stellar mass ({\mstar}), {\muv}, or the gas-phase metallicities ($Z$) derived by Rowland et al. (in prep.).  The Calzetti-like (SMC) values of $\delta=0$ ($\delta=-0.45$) lie at either edge of the shaded grey-blue region.  The average attenuation law slopes from \citet{Shivaei2020} for low and high metallicity galaxies at $z=1.4-2.6$ are shown by the grey-shaded region in the bottom right panel.  The median trend for galaxies at $z<0.3$ from \citet{Salim2018} is shown in brown and results for the $z\simeq5$ ALPINE galaxies from \citet{Boquien2022} by grey squares.  The galaxies studied in \citet{Markov2024} between $z=2-12$ are shown by the blue points and the purple contours show the results for the 3D-HST galaxies at $0.5<z<3.0$ from \citet{Nagaraj2022}.  The $\delta-A_V$ relation from radiative transfer modelling by \citet{Chevallard2013} is shown by the purple line.  Galaxies with significant ($\gtrsim4\sigma$) dust bumps have diamond markers and open markers are used for galaxies without an ALMA dust continuum detection.}
\label{delta_vs_Z_Av_M} 
\end{figure*}

In addition to the bump feature, the dependence of the slope of the dust attenuation curve on other galaxy properties can reveal information about the composition and distribution of dust in galaxies.
In Fig.~\ref{delta_vs_Z_Av_M} we plot the slope of the attenuation curve, $\delta$, against other galaxy properties.
We see a strong correlation between the dust attenuation curve slope, $\delta$, and $A_V$ (Spearman correlation coefficient $r=0.66$, p-value $=0.02$). 
No strong dependence on stellar mass, {\muv}, or gas-phase metallicity, $Z$, is seen ($|r|\leq0.13$, p-values $\geq0.68$).
We describe our results in comparison to previous studies in detail below.  

\subsubsection{Comparison to other studies}
In the top left plot of Fig.~\ref{delta_vs_Z_Av_M}, we plot $\delta$ against stellar mass and find no significant correlation. 
This is consistent with the results for a subsample of the ALPINE galaxies at $z=4.4-5.5$ from \cite{Boquien2022} and in the galaxies between $z=2-12$ from \cite{Markov2024}, which both exhibit a greater scatter in $\delta$ values.
We also compare to the results of \cite{Nagaraj2022}, who use a 5D linear interpolation model to control the dust attenuation curves for the SED fits of the 3D-HST galaxies at $0.5<z<3.0$, which also show no trend with stellar mass.  
\cite{Nagaraj2022} find that the dependence of the attenuation curve slope on mass is complex. 
Local studies such as \cite{Battisti2017}, who looked at 5500 $z<0.1$ star-forming galaxies, also find no trend.
This is in contrast to the trend seen in local star-forming galaxies at $z<0.3$ by \cite{Salim2018} who find flatter curves at higher masses.
Results from the VANDELS galaxies at redshift $z=3-4$ by \cite{Cullen2018} have also presented tentative evidence for steepening of the attenuation curve at $\log_{10}(${\mstar}/{\Msun})$~\lesssim~9.0$. 

In the top right panel, we plot $\delta$ against $A_V$ and find that the slopes of our galaxies get flatter with increasing $A_V$.
On average, the REBELS IFU attenuation law slopes are flatter than seen in local galaxies and the points are slightly offset from some of the lower redshift trends shown, such as the median relation from \cite{Salim2018} and the results from \cite{Boquien2022}.
The 23 galaxies from \cite{Boquien2022} reach low $A_V$ values even though the selection required them to have an ALMA dust continuum or {\cii}$158 \mu$m detection, which means that for a given stellar mass, the sample is biased towards the more dust-rich galaxies within the ALPINE sample.
Similar trends to these are also seen for the COSMOS galaxies at $0.1<z\lesssim3$ by \cite{Battisti2020} and the average of all the $z=2-12$ galaxies from \cite{Markov2024}.
We also compare to the results from the 3D-HST galaxies \citep{Nagaraj2022} which show a flatter trend between $\delta$ and $A_V$ that is more consistent with the position of the REBELS IFU galaxies and the $z\simeq7$ galaxies in the \cite{Markov2024} sample.
\cite{Nagaraj2022} attribute the flattening of the relation they see compared to other literature results to be due to the UV spectrum being dominated by stars that have older ages and redder colours on average and/or that are surrounded by a reduced amount of birth cloud dust in their modelling.
The \cite{Nagaraj2022} sample is also mass-complete and is therefore arguably the least biased of the studies shown. 

The strong correlation between the slope of the dust attenuation law with $A_V$ seen in observations can be reproduced using radiative transfer models.
The trend is thought to be consistent with larger effective dust optical depths and radiative transfer effects that cause the dust attenuation curve to become shallower. 
The relationship between dust attenuation optical depth and attenuation curve slope has been derived by both analytical modelling \citep{Bruzual1988} and numerically by radiative transfer codes \citep[e.g.][]{Witt1992, Witt2000, Gordon2001, Chevallard2013}.
For example, the models of \cite{Seon2016} and \cite{Inoue2006}, which include a clumpy ISM structure, predict that the steepest curves occur when the column density of dust is lowest.  
\cite{Inoue2005} also find that the amount of dust in young star-forming regions compared to the diffuse ISM (differential reddening) also plays a key role in determining the slope of the attenuation law.
We find that $A_V$ and $\delta$ correlate with the $100$~Myr SFR and it has been suggested that optical depth tends to increase with SFR.
This can be explained by the more massive stars on average facing more dust attenuation at higher SFRs since the birth clouds have an increased dust enrichment compared to the diffuse ISM \citep{Nagaraj2022}.
Furthermore, more recent work has shown that variations in the observed attenuation laws are seen even in simulations of individual molecular clouds depending on the line of sight or age \citep{DiMascia2024}.

The fact that our $\delta$ values and the $z\simeq7$ galaxies from \cite{Markov2024} are consistently shifted towards shallower slopes compared to the lower redshift trends is in line with the redshift evolution of the dust attenuation law slope found in \cite{Markov2024}.
However, our slopes are still generally steeper than the average curve at these redshifts from \cite{Markov2024}.
We find no evidence for very steep slopes ($\delta<-0.5$) as found in \cite{Boquien2022}, although these preferentially occur at $A_V\lesssim0.1$, where we only have one source.  

We find that the attenuation law slope shows no clear trend with {\muv}, although the REBELS galaxies are biased towards being UV-bright. 
This is consistent with the lack of correlation between $\delta$ and stellar mass since we expect brighter galaxies to be more massive according to the {\muv}$-${\mstar} relation \citep[e.g.][]{Duncan2014}.  

In the bottom right panel, we use the metallicities derived from the optical emission line ratios using the same NIRSpec spectra (Rowland et al. in prep.).
The REBELS IFU sample covers a range of metallicities, but no significant trend is seen.
The lack of significant correlation of $\delta$ with metallicity is consistent with the results of \cite{Salim2018} and \cite{Battisti2017}.  
The 3D-HST galaxies from \cite{Nagaraj2022} and the galaxies from \cite{Markov2024}, which mainly reside at the higher and lower metallicity end of our sample respectively, also show no clear trends. 
This is in contrast to results at $z=1.4-2.6$ from \cite{Shivaei2020} that found the average attenuation law for higher metallicity galaxies ($12+\log($O/H$)>8.5)$ to be more Calzetti-like with evidence of a bump, while for lower metallicity galaxies ($8.2<12+\log($O/H$)<8.5$) the curve was found to be steeper and exhibit no significant bump. 
This would imply that dust grains have different properties, with younger, lower-metallicity environments containing smaller grains formed from the re-processing of larger grains by ionising radiation.  
The fact a trend is seen in \cite{Shivaei2020} but not in other work could be a result of selection biases, the narrower metallicity range probed, or the fact that the grouping of the galaxies by metallicity effectively groups the galaxies by stellar mass whilst other studies consider each galaxy individually.
Interestingly, the galaxy in our sample with the lowest metallicity, REBELS-15, has a slope consistent with the Calzetti-like value and a significant ($7.2\sigma$) bump detection.  This is the opposite trend to \cite{Shivaei2020}.
The fact we and the majority of other studies see no trend could be consistent with the theory that geometry and orientation effects play more of a role in the attenuation curve shape than the optical properties of the dust grains themselves in high-redshift galaxies \citep[][Sommovigo et al. in prep.]{Chevallard2013}.

\subsection{\emph{$A_V-M_{\star}$} relation}

\begin{figure} 
\includegraphics[width=\columnwidth]{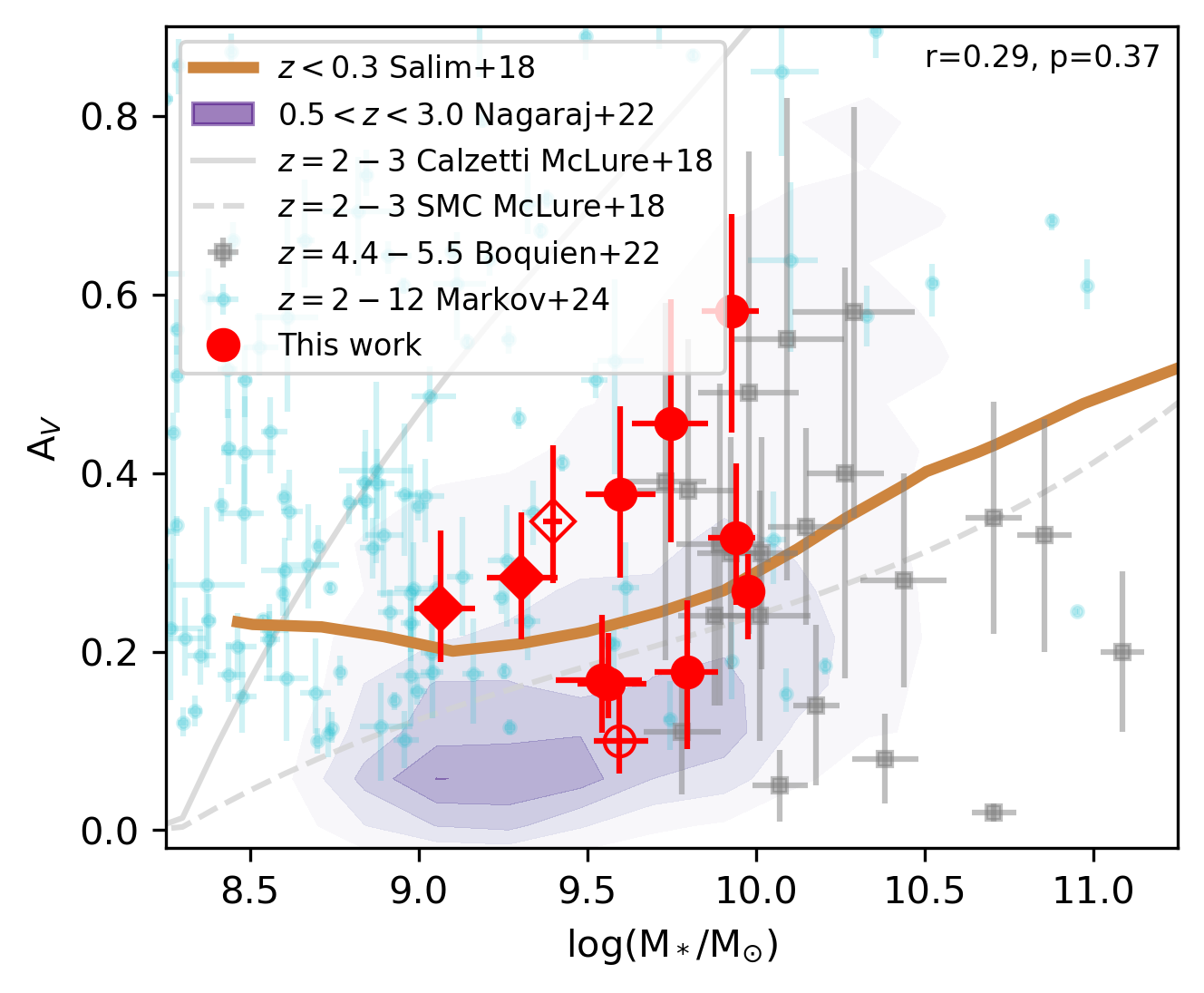}
\caption{V-band attenuation, $A_V$, versus stellar mass, {\mstar}, values for the REBELS IFU sample are shown in red.  Galaxies with significant ($\gtrsim4\sigma$) dust bumps have diamond markers and open markers are used for galaxies without an ALMA dust continuum detection. We see a weak positive correlation consistent with trends seen in local galaxies. For example, the median trend for galaxies at $z<0.3$ is shown in brown \citep{Salim2018} and the results for the 3D-HST galaxies at $0.5<z<3.0$ are shown with purple contours \citep{Nagaraj2022}.  The empirical relations for $2<z<3$ galaxies derived by \citet{McLure2018} assuming a Calzetti (SMC) attenuation law are shown by the solid (dashed) grey lines.  Results for a subsample of the ALPINE galaxies at $z=4.4-5.5$ from \citet{Boquien2022} are shown by grey squares and the galaxies at $z=2-12$ from \citet{Markov2024} are shown by blue circles.}
\label{Av_vs_mass} 
\end{figure}

The total effect of dust on the observed galaxy spectra depends on the shape of the attenuation law and the total dust attenuation, which is quantified by the value of $A_V$.
Plotted in Fig.~\ref{Av_vs_mass} are the $A_V$ values we derive as a function of stellar mass.
The $A_V$ values range between $0.10-0.58$, indicating there is significant attenuation by dust in these sources, and there is a weak positive correlation with stellar mass ($r=0.29$, p-value $=0.37$).
The recovered $A_V$ values are consistent with previous SED fitting to rest-UV photometry \citep{Inami2022} and also show a positive correlation with the dust masses derived from {\cii}in \cite{Sommovigo2022} that range between $\log(M_{\text{dust}}/${\Msun}$)~=~6.95-7.55$.

The REBELS IFU sample fills in the parameter space between the mainly lower-mass galaxies from \cite{Markov2024} and the 23 higher-mass galaxies selected by \cite{Boquien2022} from the ALPINE survey at $z=4.4-5.5$.
The $A_V$ values from these two studies were also derived from SED fitting with a flexible dust law.
Many of the REBELS IFU galaxies lie close to the median trend derived from SED fitting with a flexible dust law for local galaxies at $z<0.3$ by \cite{Salim2018}. 
The positions of the REBELS IFU galaxies also agree well with the area of the plot occupied by the 3D-HST galaxies from \cite{Nagaraj2022}. 
The shape of the $A_V$-{\mstar} relation can be used to infer the slope of the dust attenuation curve with certain assumptions. 
The majority of our galaxies sit between the two grey curves derived from the UV-continuum slopes of 8407 galaxies as a function of stellar mass at $2<z<3$ spanning a mass range of $8.5<\log(${\mstar}/{\Msun}$)<11.5$ by \cite{McLure2018} assuming a Calzetti (solid line) or SMC (dashed line) and intrinsic UV-slope of $\beta_0=-2.3$, roughly consistent with their attenuation law slopes. 
The consistency of the positions of the points with the trends from other work suggests that the relation between these parameters in star-forming galaxies does not significantly evolve with redshift.

\section{Discussion}
\label{sec:discussion}
We have derived the dust attenuation curves from \textit{JWST} NIRSpec spectra of 12 Lyman-break galaxies at $z=6.5-7.7$.
This is the first time dust attenuation curves have been measured from the rest-UV to optical spectra for these massive galaxies at $z\simeq7$.
We will discuss the features of these curves and the implications of the dependence of the dust law slope on physical properties in Sections~\ref{sec:discussion_5.1} -- \ref{sec:discussion_5.2}.
In Section~\ref{sec:discussion_5.3} we briefly discuss our results in the context of previous work using ALMA observations of these sources.
We investigate the impact of fitting using flexible dust laws on other physical parameters derived from SED fits compared to assuming a fixed Calzetti-like dust attenuation law in Section~\ref{sec:impact_assume_law} and measure the intrinsic UV-slopes of the galaxies in Section~\ref{sec:intrinsic_spectra}.

\subsection{Dust attenuation curves in massive \emph{$z\simeq7$} galaxies}
\label{sec:discussion_5.1}
The positions of the $A_V-${\mstar} points for the REBELS IFU galaxies in Fig.~\ref{Av_vs_mass} are indicative of moderate reddening and are roughly consistent with the relation seen in local galaxies, with more massive galaxies having greater dust attenuations.
Similarly, we find that $\delta$ correlates with $A_V$ but not with metallicity (Fig.~\ref{delta_vs_Z_Av_M}), which is also consistent with trends in local galaxies. 
This suggests that evolved galaxies at $z\simeq7$ have similar dust attenuation properties to galaxies in the local Universe.
Further support for this conclusion comes from studies in the FIR that find no clear evolution with redshift of the dust emissivity index, $\beta_{\text{IR}}$ \citep[e.g.][]{DaCunha2021, Bendo2023, Witstok2023b, Tripodi2024, Algera2024}.

Our dust attenuation curves are generally flatter than local sources with similar stellar masses, with no evidence for very steep slopes ($\delta<-0.5$) or significant shape deviations from local relations. 
In all but one case the dust attenuation laws shown in Fig.~\ref{BAGPIPES_dust_law_fits} lie between the Calzetti-like and SMC relations.
The majority are more consistent with the Calzetti-like relation than the steeper SMC curve.
This is consistent with previous indirect constraints placed on the dust attenuation laws of the REBELS galaxies using ALMA observations of the FIR SED and the {\irxb} relation \citep{Schouws2022, Bowler2023}.
This is also consistent with other results for galaxies with high stellar masses \citep[$\log(${\mstar}/{\Msun}$)>9$ e.g.][]{Cullen2018, McLure2018}.
Other studies using NIRSpec spectra at high redshift such as \cite{Markov2024} present tentative evidence that attenuation curves get flatter with increasing redshift from $z\simeq2-12$.
This is interpreted as being due to larger dust grains at earlier times.
However, the REBELS attenuation curves we derive are generally steeper than the median curve for the corresponding redshift range in this study. 
This could suggest that the evolution of the attenuation curve slope is not as strong as \cite{Markov2024} suggest due to their bias towards slightly lower mass galaxies.
This is supported by \cite{Nagaraj2022} who find that the redshift evolution of the dust attenuation curve is strongest for galaxies with lower masses and SFRs.

Three of our galaxies ($25$ per cent) exhibit $\gtrsim4\sigma$ evidence for a 2175~{\AA} dust bump.
This suggests small carbonaceous dust grains could be present in a significant fraction of Epoch of Reionisation galaxies.  
We note that the bump strengths in our attenuation curves provide lower limits on the bump strengths in the \emph{extinction} curves of these galaxies \citep{Salim2020}.
Evidence for dust bumps has been found in the spectra of several Epoch of Reionisation galaxies \citep{Witstok2023, Markov2023, Markov2024}.
This is surprising given that bumps are thought to be less common at early epochs due to the time needed for carbonaceous grains to build up.  
\cite{Witstok2023} find evidence for dust bumps in 10 out of 49 ($\simeq 20$ per cent) sources over $z=4.02-7.20$, with the galaxies exhibiting bumps having considerable dust attenuation (the stack of these sources had a nebular extinction derived from the Balmer decrement of $E(B-V)=0.33\pm0.01$ mag) and slightly elevated metallicities compared to the rest of the sample.  
Our sample is too small to robustly determine if there is a slightly elevated prevalence of the bump due to our galaxies probing the massive, dust-rich end of the population at these redshifts.  Other studies suggest bump strengths are correlated with metallicity, age, and stellar mass \citep[e.g.][]{Noll2009, Shivaei2020, Shivaei2022} as well as the steepness of the dust attenuation curve \citep[e.g.][]{Kriek2013, Seon2016, Narayanan2018}.
However, we see no clear trends within our sample.

\subsection{The impact of geometry on attenuation law slope}
\label{sec:discussion_5.2}
The steepness of the dust attenuation curve can be affected by the properties of the dust itself such as the grain size distribution \citep[e.g.][]{Hirashita2020, Langeroodi2024}, the dust distribution relative to the stars in the ISM, and higher concentrations of dust in star-forming regions (differential reddening).  
Radiative transfer models predict a broad range of attenuation curve shapes based on the variation of these properties \citep[e.g.][]{Seon2016}.
The steepening of a dust attenuation curve can be explained physically by the fact that red light is scattered more isotropically at low optical depths compared to blue light, which experiences more forward scattering and therefore more absorption \citep{Leja2017}. 
Shallower curves are produced at high optical depths since more of the observed light originates from outside the equatorial plane of the galaxy and is, therefore, less affected by the wavelength-dependent scattering differences \citep{Nagaraj2022}.

The relative geometric distribution of dust and stars strongly affects the attenuation curve shape \citep[][]{Salim2020}.  
In particular, the studies of \cite{DiMascia2024} and Sommovigo et al. (in prep.) show that a changing line of sight can strongly change the perceived attenuation in the UV and FUV regime due to geometric effects. 
This is indirectly supported by \cite{Cochrane2024}, who showed that orientation effects may be responsible for some "HST-dark" galaxies. 
The correlation between the slope of the dust attenuation laws with $A_V$ but not with stellar mass, {\muv}, or metallicity seen in Fig.~\ref{delta_vs_Z_Av_M} suggests that geometry is the dominant factor affecting the attenuation curves in our sample. 
The more irregular rest-UV/optical morphologies seen at higher redshifts suggest the role of geometry may become even more dominant at these redshifts \citep[e.g.][]{Huertas-Company2016, Huertas-Company2024, Faisst2017}.
From the improved spatial resolution of the IFU compared to ground-based VISTA imaging of the REBELS galaxies, it is evident that these massive galaxies have complex geometries and are often formed of multiple clumps \citep[see Figure 1 of Rowland et al. in prep.,][]{Bowler2022}.
This was predicted by numerical simulations \citep{Kohandel2020, Pallottini2022} and it is also seen in other bright galaxies at $z\simeq7$ and at lower redshifts \citep[e.g.][]{Guo2015, Barisic2017, Bowler2017, Lines2024}.  
Indeed, modelling by \cite{Witt2000} found that the trend of an increasingly grey attenuation curve with increasing dust column density was stronger in models with clumpy dust embedded in the galaxy.

The variations we see in the slopes of the attenuation curves are likely the result of a combination of effects.
For example, unobscured stellar populations may flatten the curves while scattering and different optical depths towards stellar populations of different ages may steepen the curve \citep{Lin2021}.  
Modelling by \cite{Boquien2022} showed that flatter attenuation curves are obtained at fixed $A_V$ if the stellar distribution is more extended than the dust.  
This is explained physically by a higher fraction of stars being located at the edge of the clouds where they experience lower optical depths, particularly at shorter wavelengths. 
This could be a contributing factor for REBELS-25 where we see rest-UV clumps extending beyond the ALMA dust detection \citep{Rowland2024}, probably due to differential dust obscuration since high spatial resolution dust continuum detections exhibit very different morphologies to the rest-UV.  
\cite{Boquien2022} also find that geometry plays a more important role than the exact model they use for their dust grain extinction law.  

Although we have been able to constrain the shape of the dust attenuation laws, the information this reveals about dust production mechanisms is limited.
Models, such as those of \cite{Seon2016}, have shown that the shape of attenuation curves is primarily determined by the wavelength dependence of the absorption, not the underlying extinction curve, meaning that the derived attenuation curve of the galaxy cannot uniquely constrain the extinction curve. 
In agreement with this, models by \cite{Lin2021} have shown that a wide range of extinction curves can produce similar attenuation curves.
However, this modelling also shows that additional constraints may be possible in the future using ALMA observations of the FIR SED and the {\irxb} relation.
This highlights the value of the multi-wavelength observations available for the REBELS galaxies.

\subsection{Comparison to FIR observations}
\label{sec:discussion_5.3}
While a full comparison to FIR properties is beyond the scope of this paper, we discuss briefly here the ALMA properties of the sources.
When we compare to previous work based on the ALMA observations of these galaxies we see sensible trends such as blue ($\beta<-2$) UV-slopes in the two galaxies (REBELS-15 and REBELS-34) which show no dust continuum detections from ALMA \citep{Inami2022}.
REBELS-34 also has the lowest $A_V$ value. 
Consistent with what we would expect from local relations (Section~\ref{sec:results_4.2}), larger $A_V$ and $\delta$ values are seen in galaxies with higher FIR to UV luminosity ratios \citep[{\irx}; ][]{Bowler2023, Algera2024}. 
Similarly, the dust masses derived from {\cii}presented in \cite{Sommovigo2022} \citep[and][for REBELS-25]{Algera2024} correlate positively with both $A_V$ and $\delta$, except for REBELS-25 which is an outlier. 
This implies that including the ALMA data in the SED fitting may be important for sources with high FIR luminosities and large obscured star-formation fractions like REBELS-25. 
Since the dust is spatially offset from the rest-UV emission in REBELS-25 \citep{Rowland2024} its observed spectrum could be dominated by the UV-luminous regions with lower dust obscuration, causing $A_V$ to be underestimated.
Spatially-resolved SED fitting is needed to investigate this further.

Comparison with quantities derived from the rest-frame FIR observations also supports the physical interpretation of our results.
If the flattening of attenuation curves seen in our sample is due to radiative transfer effects in a clumpy medium, we should see flatter slopes in galaxies with a higher molecular index, $I_m = (F_{158}/F_{1500})/(\beta-\beta_0)$, where $F_{158}$ is the observed FIR continuum flux at $\lambda_{\text{rest}} \approx 158 \mu$m and  $F_{1500}$ is the observed UV-continuum flux at $\lambda_{\text{rest}} = 1500$ {\AA} \citep[see][for details]{Ferrara2022, Inami2022}.
Generally, the galaxies with shallower dust attenuation curves have higher molecular indices, which supports our geometrical interpretation.
For example, REBELS-38 has the flattest slope and the highest $I_m=1787$.
However, the exception to this trend is REBELS-25, which has a similar $I_m=1772$ value suggesting it is equally clumpy but it exhibits a steeper attenuation curve slope.
This could suggest that, while geometry seems to dominate in the majority of our sample, other factors such as the changing size distribution of grains may also be important in some sources since the shape of the attenuation curve depends on a complex combination of effects.
However, we caveat again that the spectrum of REBELS-25 may be dominated by regions of lower dust obscuration.

Finally, we also find that $A_V$ and $\delta$ correlate with the spatial offsets between the UV and dust continuum calculated in \cite{Inami2022}.
Again, this supports that geometric effects play an important role in our attenuation curves since these UV-dust offsets imply different star-forming regions have different dust obscurations.
Future work utilising recently obtained high spatial resolution ALMA dust continuum observations is needed to understand the dust properties of these sources in greater detail and will be the subject of a future paper (Fisher et al. in prep.).  Better constraints on systematics such as dust temperatures from ongoing observations will also improve combined \textit{JWST}-ALMA analysis.

\subsection{Impact of assuming dust laws on galaxy properties}
\label{sec:impact_assume_law}
\begin{figure} 
\includegraphics[width=\columnwidth]{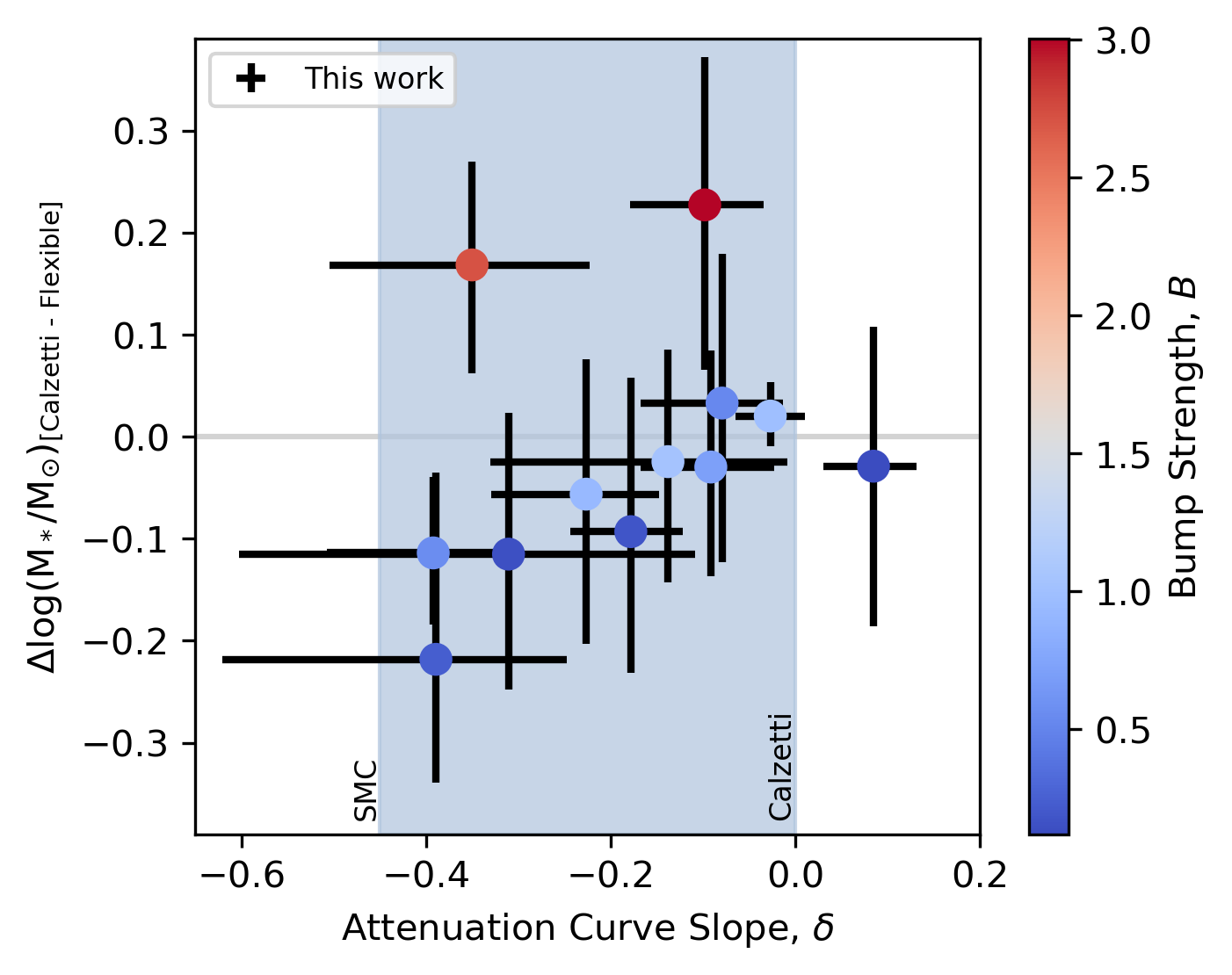}
\caption{Stellar mass difference between SED fits assuming a Calzetti-like dust attenuation law and with a flexible dust attenuation law, shown against the best-fitting dust attenuation law slope.  The points are coloured by the strength of the 2175~{\AA} dust bump. The difference in stellar mass correlates with the deviation of the dust law slope from the Calzetti-like relation except for galaxies with a strong dust bump.  In these two cases, the assumption of a Calzetti-like attenuation curve results in their masses being overestimated by up to $0.3$~dex.}
\label{Difference_vs_Cal} 
\end{figure}

Since assuming a standard dust law template is common in SED fitting, we investigated the impact assuming a Calzetti-like dust attenuation law has on the derived galaxy properties. 
In Fig.~\ref{Difference_vs_Cal} we show the differences in the stellar mass values derived with a flexible dust law compared to SED fitting with a fixed Calzetti-like dust attenuation law. 
Unsurprisingly, the stellar mass differences are negligible, with a median of $0.07$~dex, owing to the similarity in the slope of the dust laws to Calzetti and correlate with $\delta$ if no strong dust bump is present.
Assuming a Calzetti-like dust attenuation law produces masses up to $0.23$~dex larger in the two galaxies with the strongest dust bumps (REBELS-08 and REBELS-25).
The true masses are lower since the bump feature results in a rest-UV flux deficit. 
If instead we fix the dust attenuation law to the SMC relation, the mass differences are slightly larger, up to a maximum of $0.41$~dex for REBELS-08.
We note that any differences in the masses used in this work to those derived from SED fits assuming a Calzetti-like dust attenuation that are presented in Stefanon et al. (in prep.), Algera et al. (in prep.), and Rowland et al. (in prep.) are not significant enough to affect the trends discussed in these studies.

The size of the mass differences are consistent with previous results for the REBELS galaxies from SED fits to photometry by \cite{Topping2022} that found masses from the fits assuming an SMC dust law were on average 0.09 dex lower than those assuming a Calzetti attenuation law.
\cite{Topping2022} also show that the differences from the assumed SFH are more significant than from the assumed dust attenuation law.
These results are also consistent with other studies that compare derived properties found using a Calzetti-like curve to a flexible attenuation law.
For example, at $z\simeq5$ \cite{Boquien2022} find the fractional change in their mass estimates is $0.08\pm0.14$ and in local ($z<0.3$) galaxies \cite{Salim2016} found that assuming a Calzetti dust law produced masses that were on average only 0.06 dex lower than with a modified attenuation law.  

Our $A_V$ values can deviate by up to $0.39$~dex and the SFR averaged over 100 Myr by up to $0.18$~dex when using a Calzetti-like curve compared to using the derived attenuation curves.
As expected, the change is less than that found in \cite{Topping2022}, who find SFRs could differ by up to $0.4$~dex when comparing results using Calzetti to SMC because the spectra provide more information than photometry.
\cite{Markov2023} found that stellar masses, SFRs, and $A_V$ values stay consistent within $1-2\sigma$ when assuming a Calzetti-like curve compared to the flexible model in three galaxies at $z=7-8$.  The larger deviations of up to $0.4$ dex when assuming an SMC or MW dust in this study are only seen in the galaxies with very extreme attenuations of $A_V>1$.  
Other parameters of our SED fits such as metallicity and $\log U$ remain consistent within the errors.
Additionally, the metallicities derived from optical emission line ratios remained consistent within the errors (Rowland et al. in prep.). 
Thus, we conclude that our dust attenuation curves make minimal difference to the derived properties of the REBELS IFU galaxies due to the moderate $A_V$ values.
Future spatially resolved analysis of these galaxies is needed to see if other effects, such as outshining \citep[e.g.][]{Gimenez-Artega2024}, significantly affect physical properties obtained from the integrated spectra.  

\subsection{Intrinsic spectra}
\label{sec:intrinsic_spectra}
\begin{figure} 
\includegraphics[width=\columnwidth]{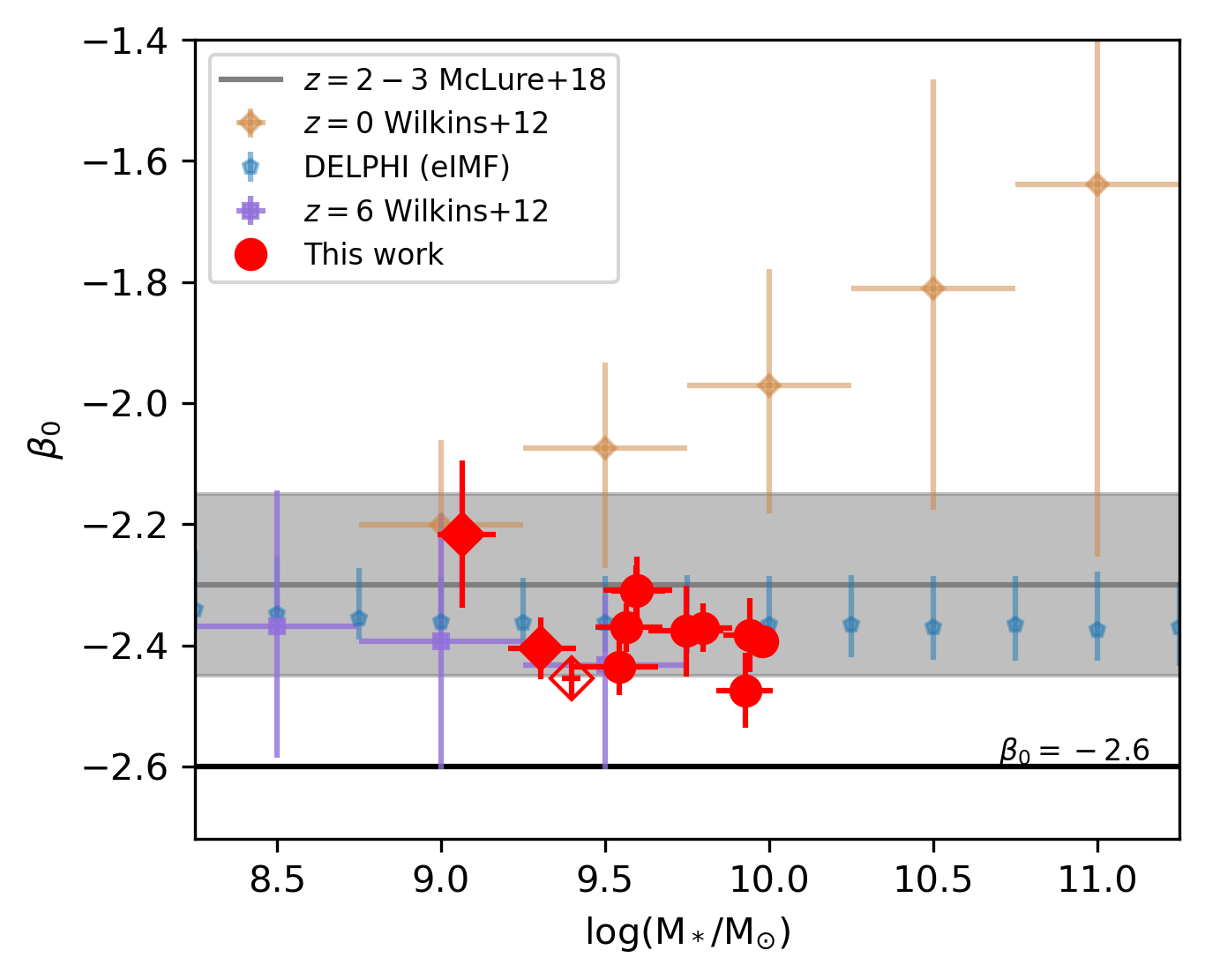}
\caption{The intrinsic UV-continuum slopes, $\beta_0$, plotted against stellar mass, {\mstar}, are shown in red.  These were obtained from applying the dust attenuation curves to the fitted SED model and have a mean value of $\beta_0=-2.38\pm0.08$ implying these galaxies contain evolved stellar populations.  The horizontal black line shows the theoretical minimum value of $\beta_0=-2.6$ \citep{Cullen2017}.  The value of $\beta_0=-2.3\pm0.15$ from SED fits of star-forming galaxies at $2<z<3$ from \citet{McLure2018} is shown in grey.  The brown diamonds and purple squares show the modelling results from \citet{Wilkins2012} for galaxies at $z=0$ and $z=6$, respectively. The blue pentagons show the intrinsic slopes from DELPHI for a model whose initial mass function varies depending on redshift and metallicity (eIMF, Mauerhofer et al. in prep.). }
\label{int_beta_vs_mass} 
\end{figure}

Applying the dust attenuation curves to the fitted SED models we obtained an intrinsic spectrum for each galaxy with the effect of dust removed.
The intrinsic UV-slope can reveal information about the recent SFH and metal enrichment history, as well as constraining the stellar initial mass function.  
The values measured for our sample shown in Fig.~\ref{int_beta_vs_mass} range between ${\beta_0=-2.22}$ and ${\beta_0=-2.47}$. 
This is in excellent agreement with results using photometry from \cite{Bowler2023} for the REBELS galaxies, who found intrinsic slopes between $\beta_0=-2.3$ and $\beta_0=-2.5$.
The intrinsic slopes also agree with the value of $\beta_0=-2.30\pm0.15$ from SED fits to $2<z<3$ star-forming galaxies of similar masses by \cite{McLure2018}.
The intrinsic UV-slope values are all greater than the theoretical minimum value of $\beta_0 = -2.6$ obtained by assuming a young ($\lesssim30$~Myr), dust-free stellar population with maximum nebular continuum emission that reddens the UV-slope \citep[e.g.][]{Cullen2017, Reddy2018}.
Thus, our results imply that the massive REBELS galaxies are more evolved with populations of less massive stars with redder intrinsic spectra already present, although some models find these slopes can be produced more quickly with higher stellar metallicities \citep{Topping2022b}.
The SED fits for the galaxies with the redder intrinsic UV-slopes also tend to have higher metallicities. 
Higher metallicities cause redder UV-slopes since the stellar absorption features in the rest-UV increase with metallicity \citep[e.g.][]{Calzetti1994, Calabro2021}.  

Simulations have suggested that redshift is the most important factor for setting $\beta_0$ at $z\geq6$ since the correlations between $\beta_0$ and galaxy properties such as the SFR, intrinsic UV luminosity, and stellar mass flatten out with redshift \citep{Wilkins2012}.
In Fig.~\ref{int_beta_vs_mass} we see a weak negative correlation with mass such that the sources with higher masses have bluer intrinsic UV-slopes. 
The $z=6$ simulation results from \cite{Wilkins2012} also exhibit a slight downward trend with mass, which is the opposite trend to that seen at $z=0$.  
Our intrinsic UV-slopes also show excellent consistency with more recent results from the semi-analytical galaxy formation code DELPHI (Mauerhofer et al. in prep.).
We show the intrinsic UV-slope from their eIMF model (but note that the values for their Fiducial model are almost identical).
This model has a variable IMF that depends on redshift and metallicity such that the IMF becomes more top-heavy at higher redshift and lower metallicities.
This allows the model to reproduce a UV luminosity function consistent with observations over a broad range of redshifts.

The range of intrinsic UV-slopes seen demonstrates the validity of concerns about assuming a fixed intrinsic UV-slope and the scatter this introduces to relations that assume a fixed value for this parameter (e.g. {\irxb}), especially given that from the selection of these galaxies we might expect the REBELS IFU sample to have similar intrinsic properties.  
This shows the value of spectroscopic observations at these redshifts and reinforces the known limitations of deriving dust attenuation laws from relations such as $A_V$--{\mstar} or {\irxb}.
While these may capture the average properties for a population of galaxies, we recommend caution when applying corrections to individual sources.

\section{Conclusions}
\label{sec:summary} 
We measure the dust attenuation laws of 12 massive ($9~<~\log$(\mstar/\Msun)~$~<~10$) Lyman-break galaxies from \textit{JWST} NIRSpec spectra at $z\simeq6.5-7.7$.  
Our key findings are: 

\begin{itemize}
    \item We find that the dust attenuation laws for individual galaxies exhibit a range of slopes.  The slopes correlate with $A_V$, but there is no clear dependence on stellar mass, {\muv}, or gas-phase metallicity.  Comparing this to empirical models suggests that the most important factor driving the steepness variation in the attenuation curves is dust-star geometry and not differences in the chemical composition or grain size distribution of the dust itself.  This is supported by the clumpy geometry revealed by the spatial resolution of the IFU and the UV-FIR offsets seen in high-resolution ALMA observations of these sources.  
    \item The mean attenuation law slope of $\delta=-0.18\pm0.15$ is consistent with the Calzetti-like ($\delta=0$) relation in local starburst galaxies.  The attenuation curves are generally flatter than local sources with no evidence for significant deviations in shape from local relations or very steep slopes of $\delta<-0.5$. We see $A_V$ values indicative of moderate reddening ($A_V=0.1-0.6$ mag) that have a weak positive correlation with stellar mass, indicating more massive galaxies are more dust-rich, as seen in local galaxies.  This suggests that evolved galaxies at $z\simeq7$ have similar dust properties to local sources.  
    \item Three dust attenuation curves ($25$ per cent of our sample) exhibit 2175~{\AA} dust bumps with bump strengths at a significance of $\gtrsim4\sigma$.  These are the most massive galaxies at $z\simeq7$ found to have this signature to date.  This suggests small carbonaceous dust grains may be present in a significant fraction of Epoch of Reionisation galaxies.  This places constraints on dust formation mechanisms since these grains must therefore form rapidly \citep{Witstok2023, Schneider2023}.
    \item The differences in derived parameters such as stellar mass and metallicity using the flexible dust attenuation law compared to assuming a Calzetti-like curve are negligible.  The mass differences correlate with the attenuation law slope, except for the galaxies with strong $2175$~{\AA} dust bumps.  Not accounting for the deficit in the rest-UV flux caused by these dust bumps can cause the stellar masses to be overestimated by up to $0.3$~dex. 
    \item The derived intrinsic UV-slopes from our fitting have a mean value of $\beta_0=-2.38\pm0.07$ and are in excellent agreement with previous results from SED fits to photometry \citep{Bowler2023}.  All are greater than the theoretical minimum value of $\beta_0=-2.6$, suggesting our galaxies contain evolved stellar populations.  We recommend caution when using a fixed value for $\beta_0$ (e.g. when correcting SFRs from the {\irxb} relation) given the range of values we obtain.
\end{itemize}

Gaining further insights into the dust properties of these sources will require utilising the spatial resolution of the IFU and complementary ALMA observations available for these sources. 
For example, we can make spatially-resolved measurements of the rest-frame UV-slope and Balmer decrement allowing us to investigate the spatial distribution of dust, gas, and stars within these sources.
These observations can also reveal to what extent effects such as outshining affect the properties measured from the integrated spectra. 
For six of the REBELS IFU galaxies high-resolution (up to $0.15''$) ALMA Band 6 observations have recently been obtained \citep[see][for REBELS-25 and Phillips et al. in prep.]{Rowland2024}. 
This will allow us to investigate how rest-UV and rest-optical features correlate with the position of the dust continuum detections, further revealing the properties of dust in the first Gyr of the Universe.

\section*{Acknowledgements}
We thank the authors of \cite{Markov2024}, \cite{Nagaraj2022}, and \cite{Salim2018} for kindly sharing their data from these papers. 
RB acknowledges support from an STFC Ernest Rutherford Fellowship [grant number ST/T003596/1].
MA acknowledges support from ANID BASAL project FB210003 and ANID MILENIO NCN2024\textunderscore112.
PD acknowledges support from the NWO grant 016.VIDI.189.162 (``ODIN") and warmly thanks the European Commission's and University of Groningen's CO-FUND Rosalind Franklin program.
AF acknowledges support from the ERC Advanced Grant INTERSTELLAR H2020/740120. 
JH acknowledges support from the ERC Consolidator Grant 101088676 
(“VOYAJ”).
N.S.S. gratefully acknowledges the support of the Research Foundation - Flanders (FWO Vlaanderen) grant 1290123N.

\section*{Data Availability}
The data used in this manuscript will be made available to others upon reasonable request to the authors.


\bibliographystyle{mnras}
\bibliography{export_16Dec_3} 




\appendix

\section{SED model tests}
\label{sec:appendix} 
This section describes the tests we performed to check that our SED fitting procedure accurately recovers the slope of the dust attenuation law.  
We used $\tt{BAGPIPES}$ to generate model galaxy SEDs with the same observed wavelength coverage as the real NIRSpec spectra for our sample.
For each of the twelve REBELS IFU galaxies, we take the best-fit parameters from our SED fits to the real spectra for all the properties except for $A_V$ and $\delta$.
We then generate eight model galaxy spectra for four values of $\delta=[0.0, -0.2, -0.4, -0.6]$ and two values of $A_V = [0.2, 0.5]$.
These are representative of the range of values that we see in our sample.
We then add realistic Gaussian noise to the model SED.
To do this we fit a second-order polynomial to the error on the observed spectrum of one of our observations to capture the shape of its wavelength dependence and then scale this according to the median rest-UV flux between rest-frame wavelengths of 1216-3500~{\AA}.
This produces spectra with noise levels that are visually similar to what we see in the real spectra.
We then use the same fitting procedure described in Section~\ref{sec:methods}.

\begin{figure*}
\includegraphics[width=\columnwidth]{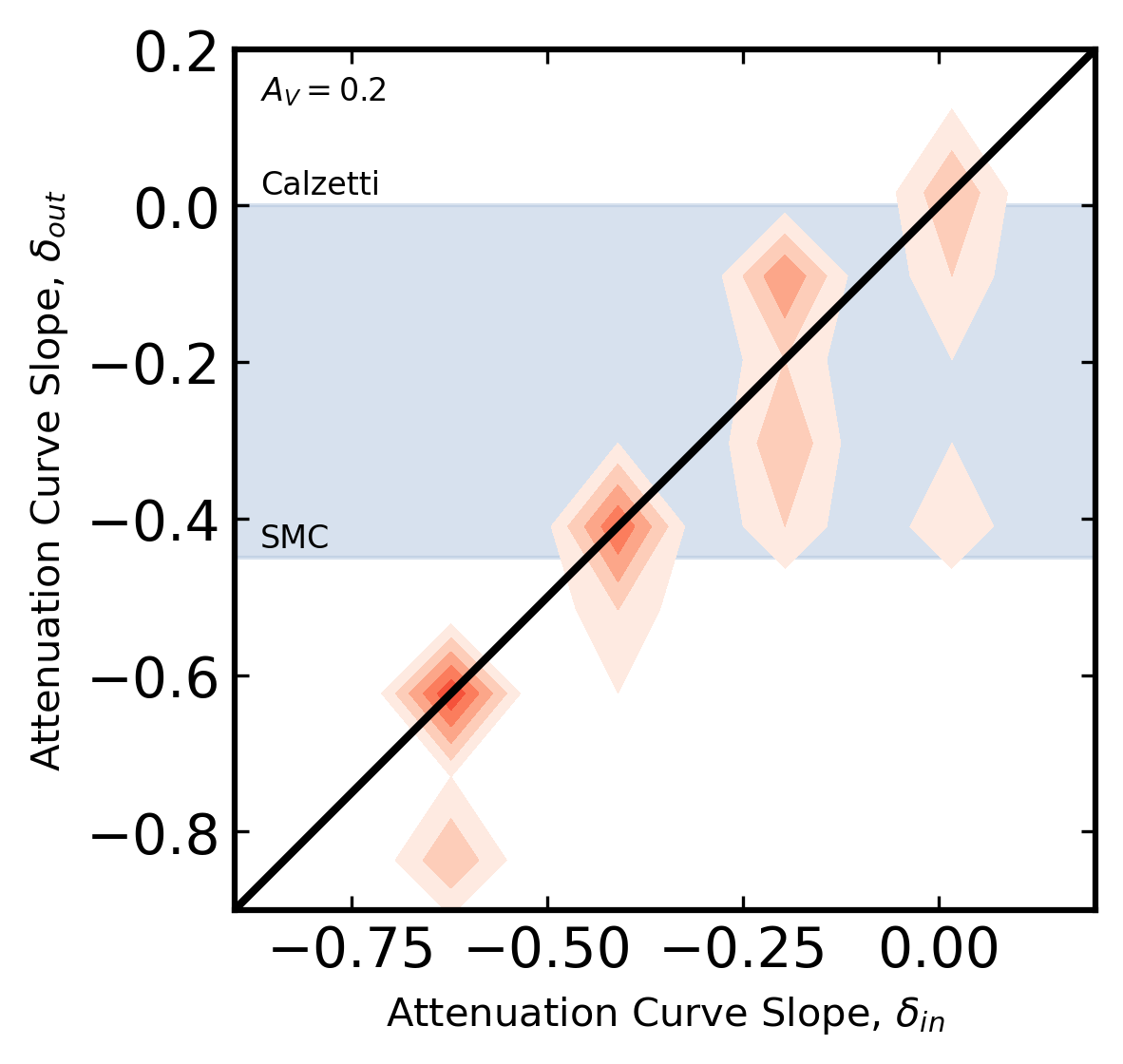}
\includegraphics[width=\columnwidth]{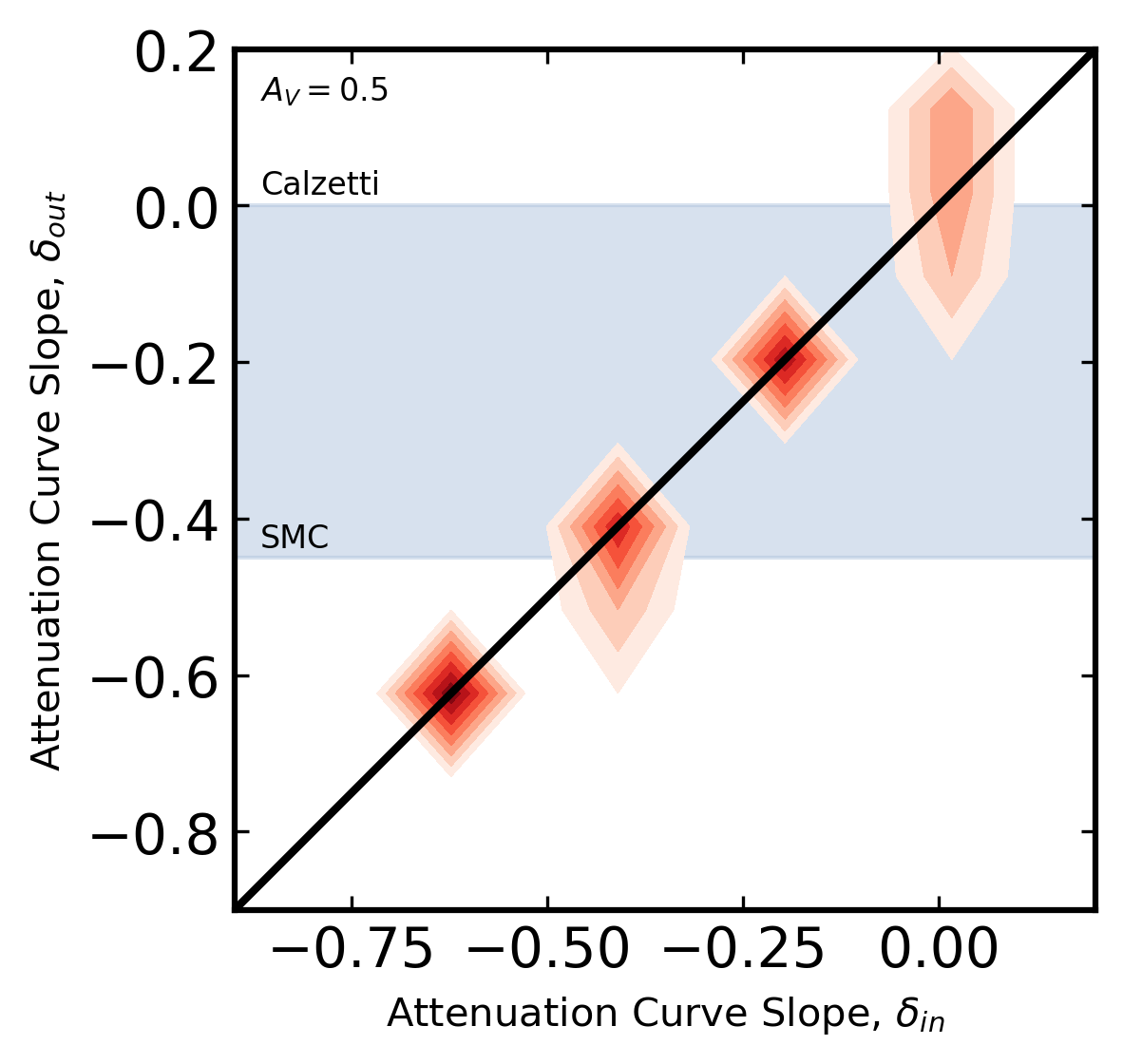}
\caption{Contour plot for the fitted dust attenuation law slope, $\delta_{out}$, plotted against the slope of the input model $\delta_{in}$. For each of our twelve galaxies, we generate eight model galaxy spectra for four values of $\delta=[0.0, -0.2, -0.4, -0.6]$ and two values of $A_V = [0.2, 0.5]$.  We see that the dust attenuation law slope is recovered well, particularly when the curve is steeper.  As expected, the scatter is also lower at higher $A_V$ since dust has a greater effect on the spectra. }
\label{SED_fit_tests} 
\end{figure*}

In Fig.~\ref{SED_fit_tests} we fit contours to the recovered dust attenuation law slope against the slope inputted into the model.  
The peak of each distribution at each $\delta$ value is close to the 1-to-1 line, confirming that we can accurately recover the attenuation law. 
Given that some of these $A_V-\delta$ scenarios may be physically unlikely, the true recoverability is expected to be better than implied here. 
By plotting the contours at the same levels in both plots we can see that the scatter in recovered slopes is lower at higher $A_V$, as we would expect since at lower $A_V$ the dust will have a smaller impact on the observed spectrum.
This is reflected in the larger error bars on $\delta$ values for sources with low $A_V$ in Fig.~\ref{delta_vs_Z_Av_M}.
The slope is also better recovered in both $A_V$ cases when it is steeper, confirming that if steep slopes were present in our sample we would recover them.  

\begin{figure*}
    \includegraphics[width=2\columnwidth]{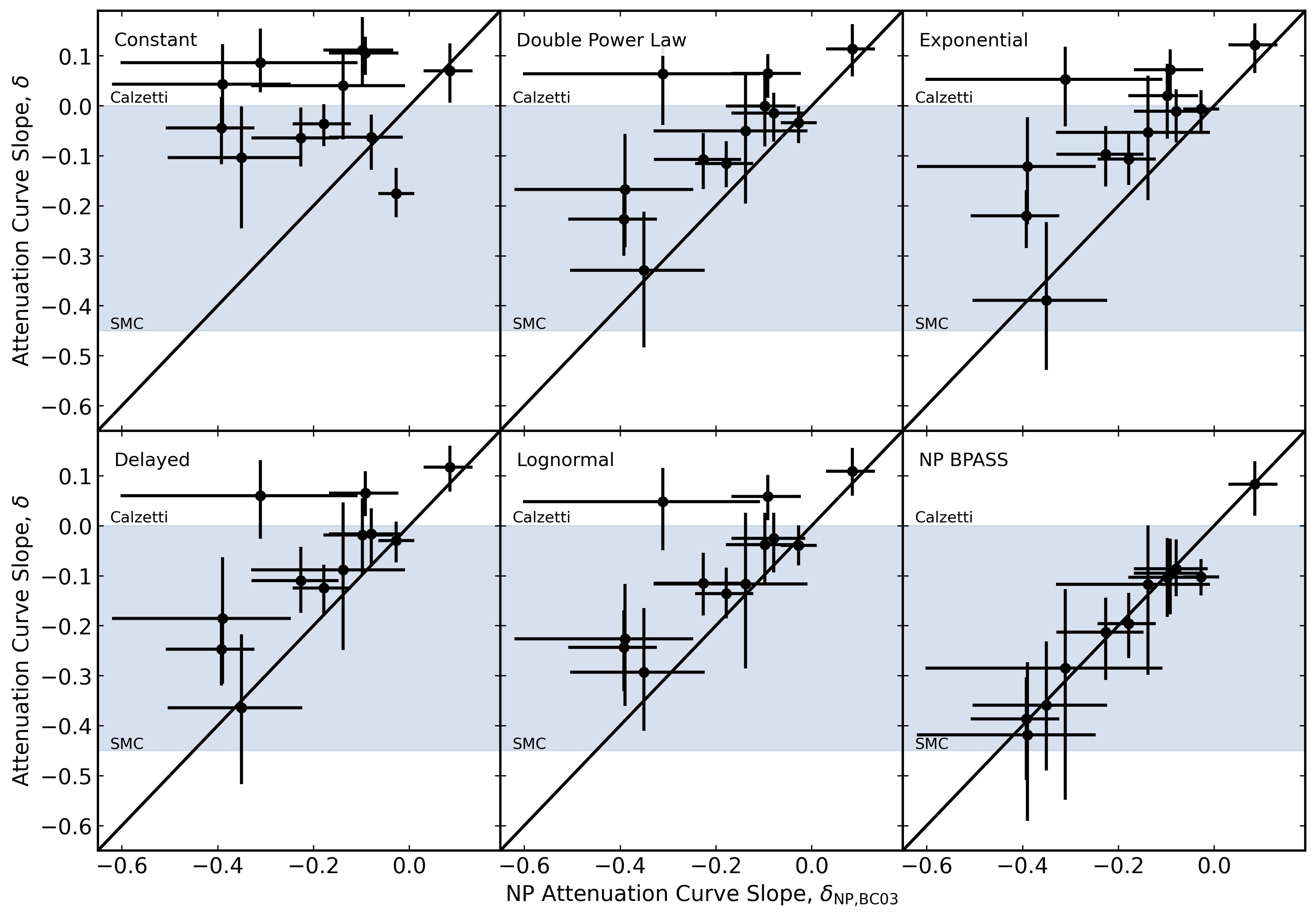}
    
    \caption{The fitted dust attenuation law slope, $\delta$, for different SFHs compared to the values used throughout the rest of this work that used a non-parametric SFH. We also show the slopes recovered when using the non-parametric SFH with the BPASS stellar population models are consistent within errors with our fiducial fits. }
    \label{SED_fit_tests_SFH} 
\end{figure*}

In Fig.~\ref{SED_fit_tests_SFH} we show the impact of assuming different SFHs on our results.
We run the $\tt{BAGPIPES}$ fits to the observed spectra with continuity, constant, double power law, exponential, delayed, and lognormal SFHs.
The largest deviations from the fiducial non-parametric results we present in the rest of this work, $\delta_{\text{NP,BC03}}$, are seen when using the constant SFH, which recovers very flat slopes.
This is unsurprising given it is the model with the least flexibility.  
However, in all cases, we recover the same trends as those shown in Fig.~\ref{delta_vs_Z_Av_M} and \ref{Av_vs_mass}.
Thus, we can be confident that our results and conclusions are not significantly affected by our choice of SFH.  
We also run using the non-parametric (NP) SFH but with the BPASS stellar population models and find the recovered slopes are consistent within the errors.  

\section{SED fits to the spectra for all 12 galaxies}
\begin{figure*}
    \begin{tabular}{ll}
    \includegraphics[width=\columnwidth]{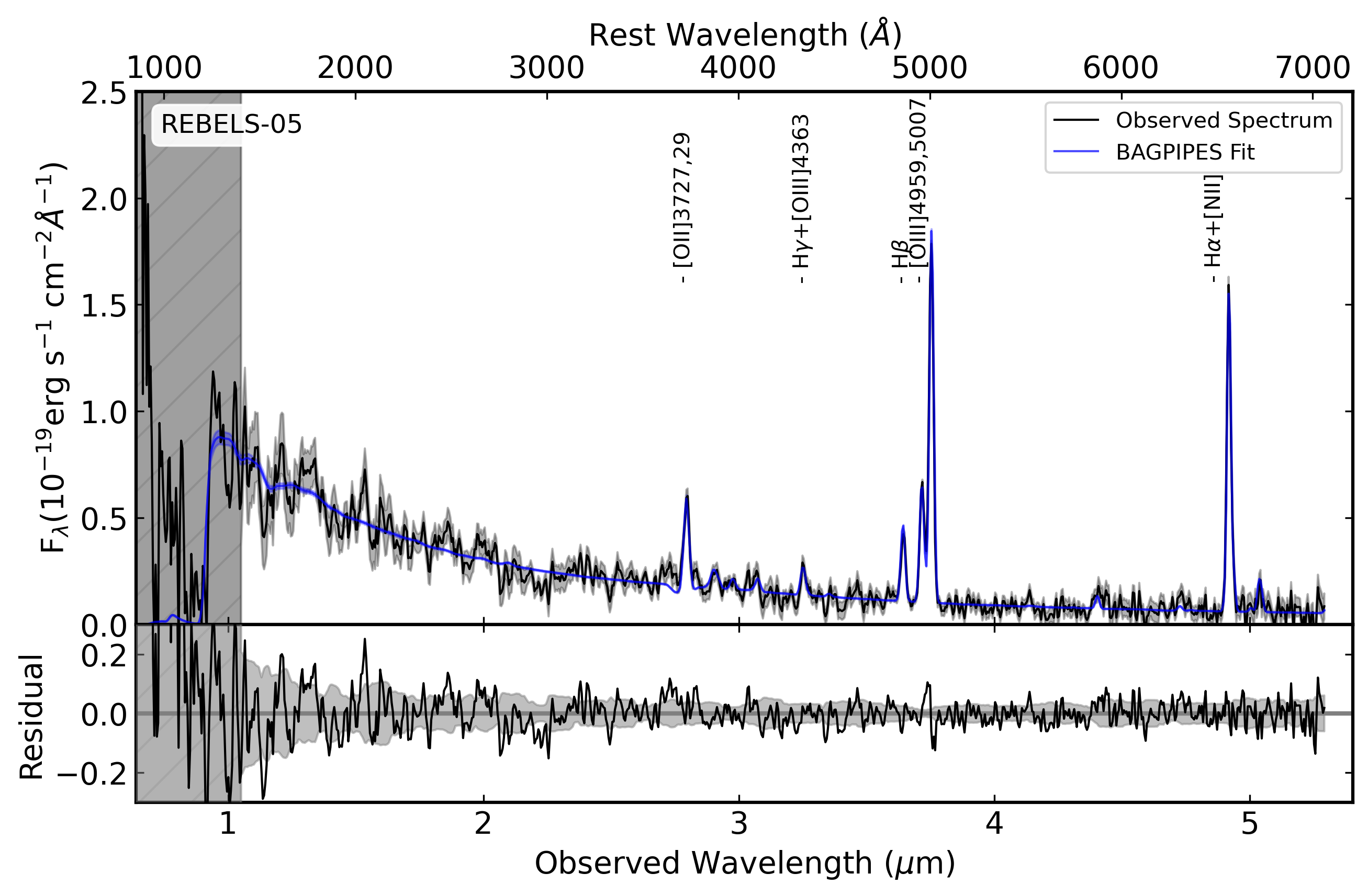} & 
    \includegraphics[width=\columnwidth]{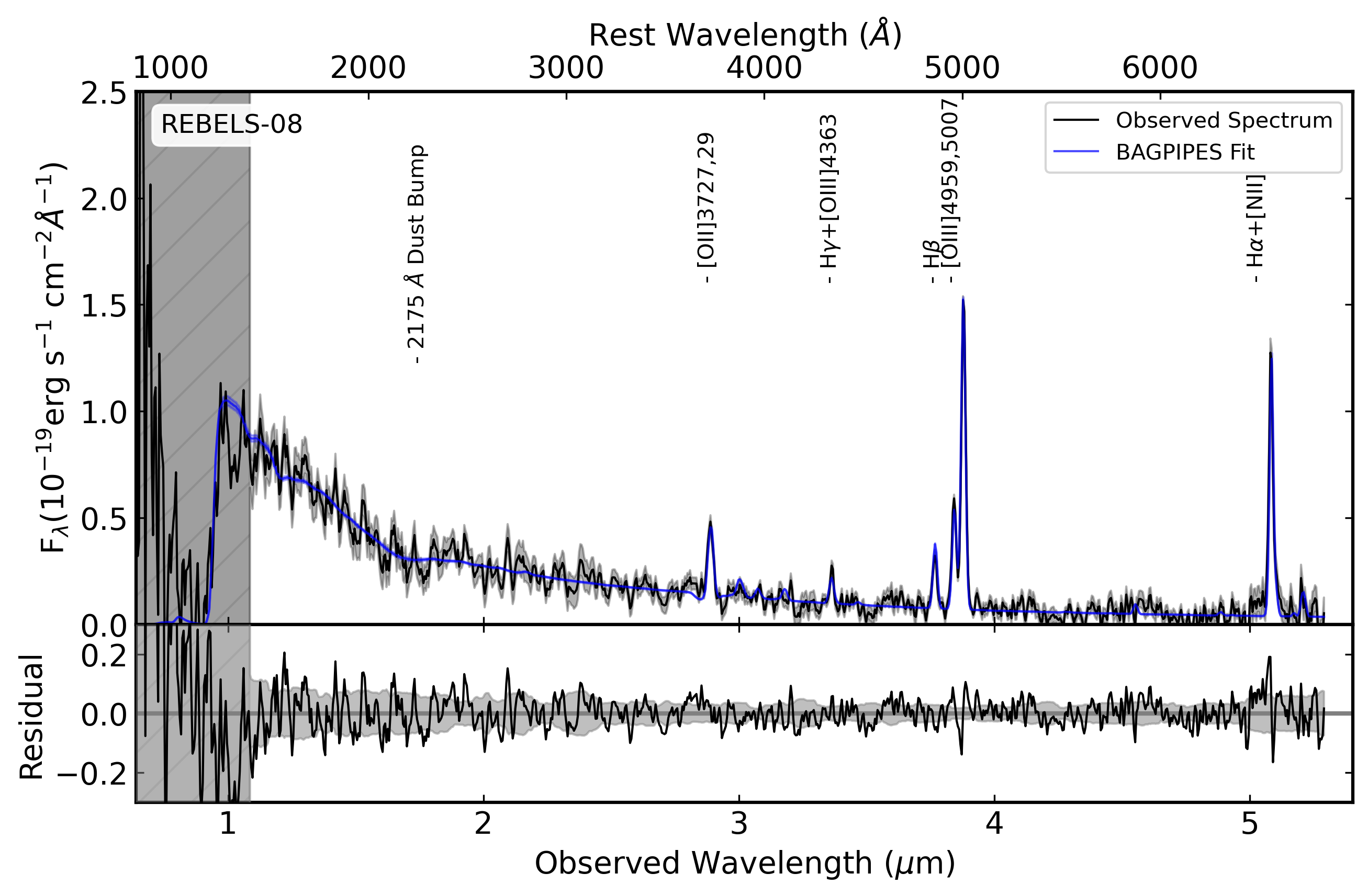} \\ 
    \includegraphics[width=\columnwidth]{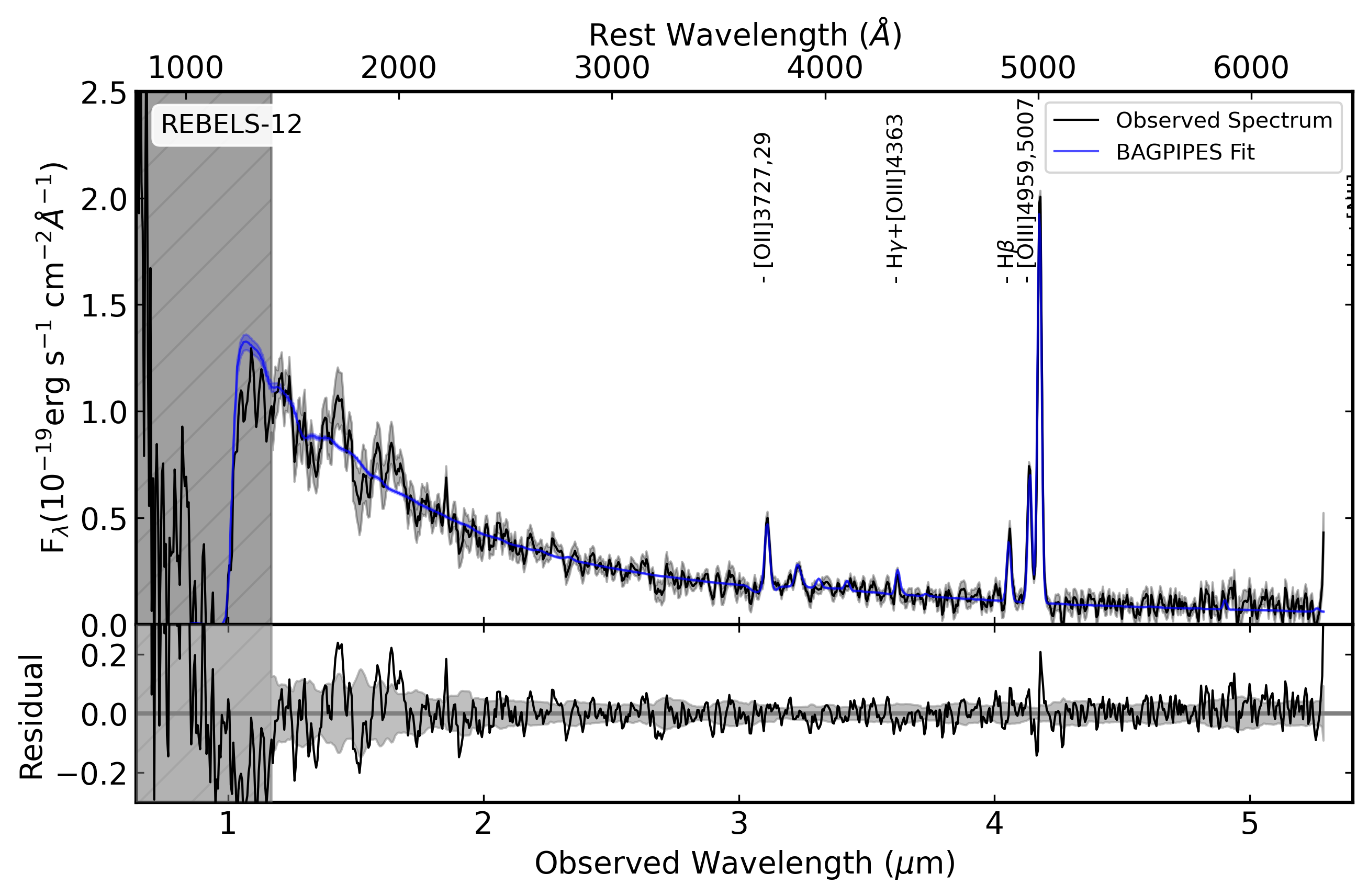} & 
    \includegraphics[width=\columnwidth]{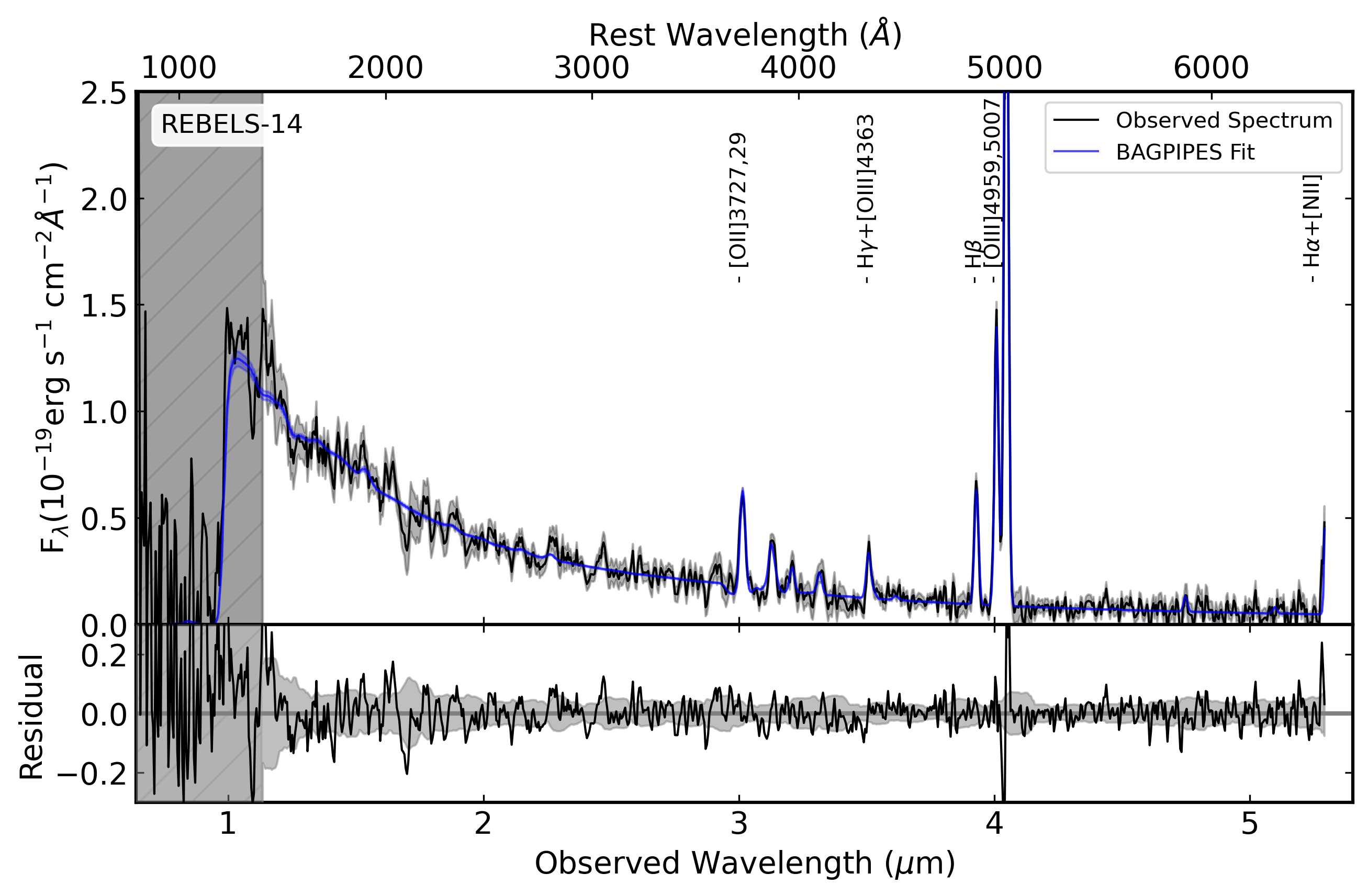} \\ 
    \includegraphics[width=\columnwidth]{REBELS-15_full_galaxy_bagpipes_fit_NP_BC03MaskDLy1400Eta.png} & 
    \includegraphics[width=\columnwidth]{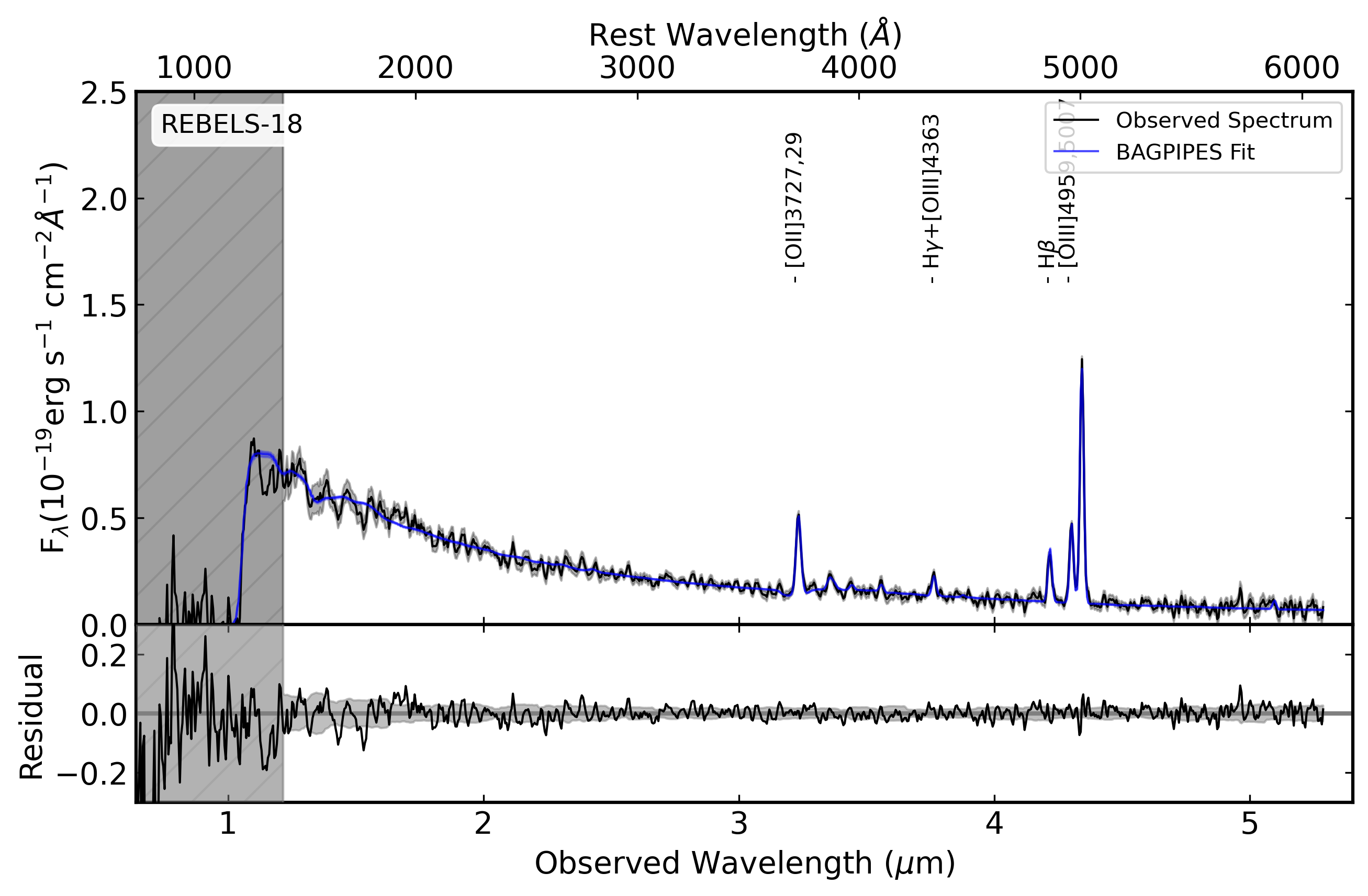} \\ 
    
    \end{tabular}
    \caption{Full NIRSpec PRISM spectra for the REBELS IFU galaxies with the 1$\sigma$ error shown by the grey shaded region.  The best-fitting SED model using a flexible dust attenuation law is shown in blue.  The spectrum is masked below a rest-frame wavelength of 1400~{\AA} to exclude effects caused by potential Ly$\alpha$ damping and instrumental effects at low wavelengths.  The bottom panel in each plot shows the fit residuals.  REBELS-08, REBELS-15, and REBELS-25 show evidence for a 2175~{\AA} dust bump.}
    \label{Spectra_appendix1} 
\end{figure*}

\begin{figure*}
    \begin{tabular}{ll}
    \includegraphics[width=\columnwidth]{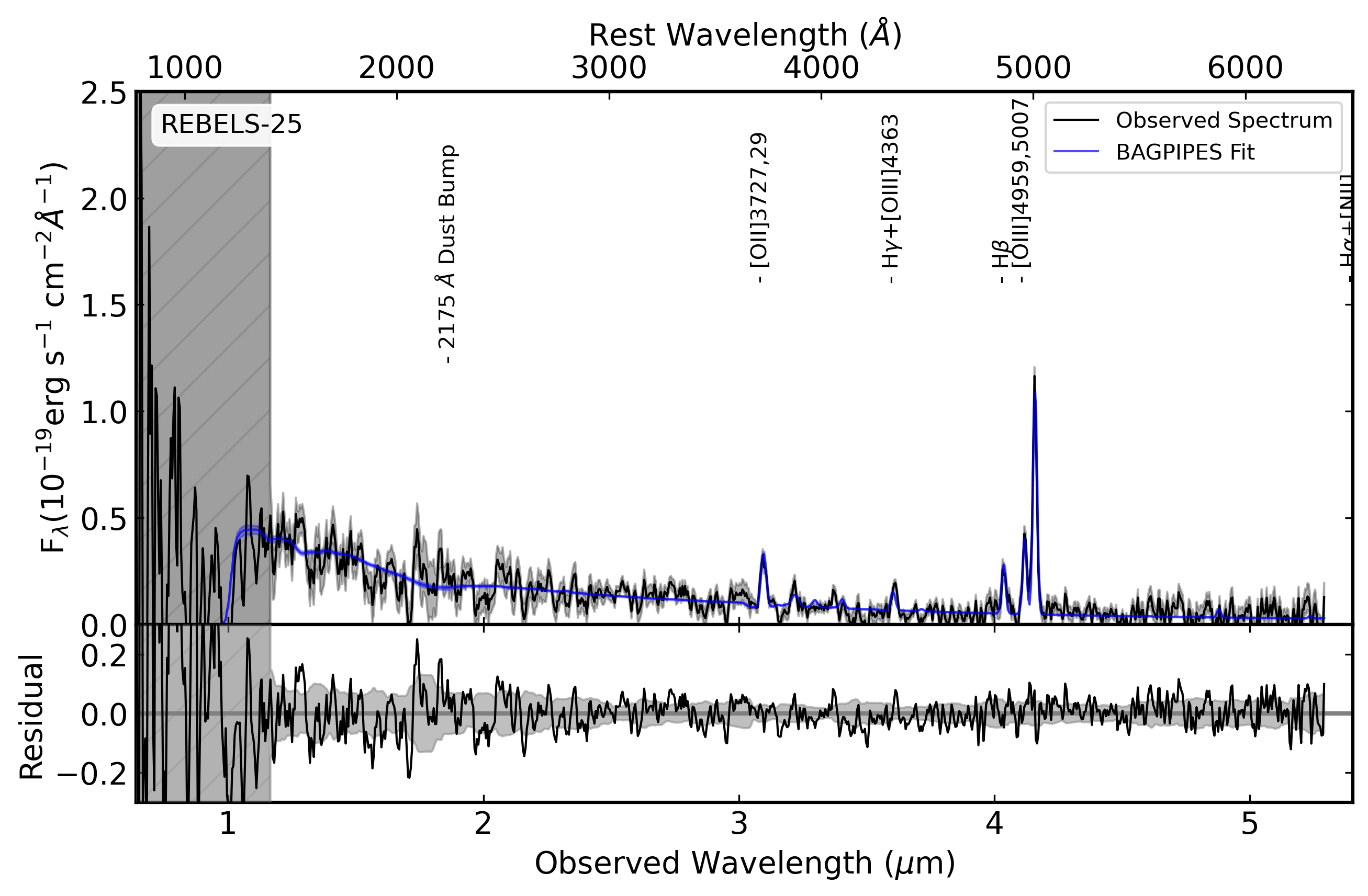} & 
    \includegraphics[width=\columnwidth]{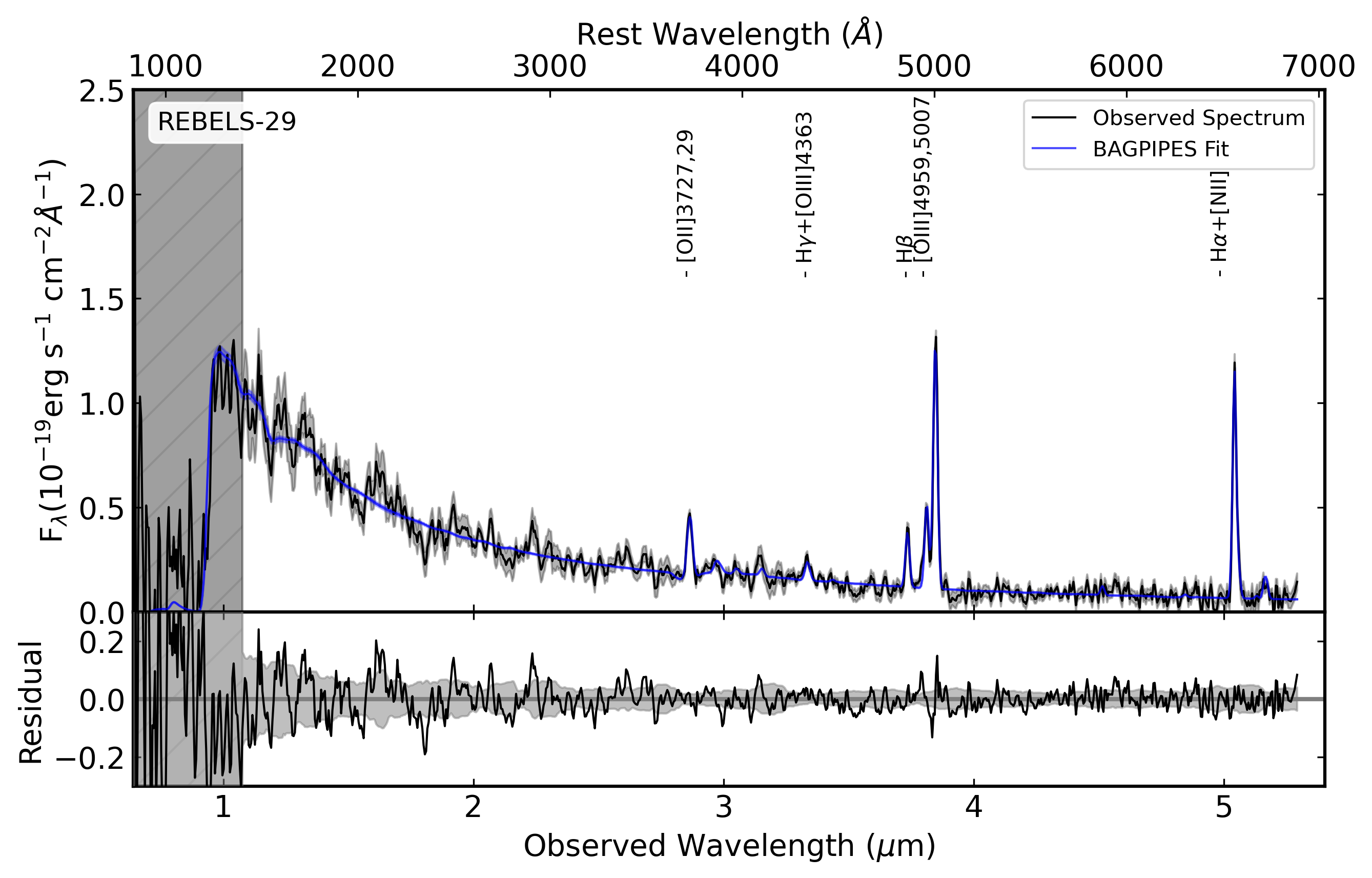} \\ 
    \includegraphics[width=\columnwidth]{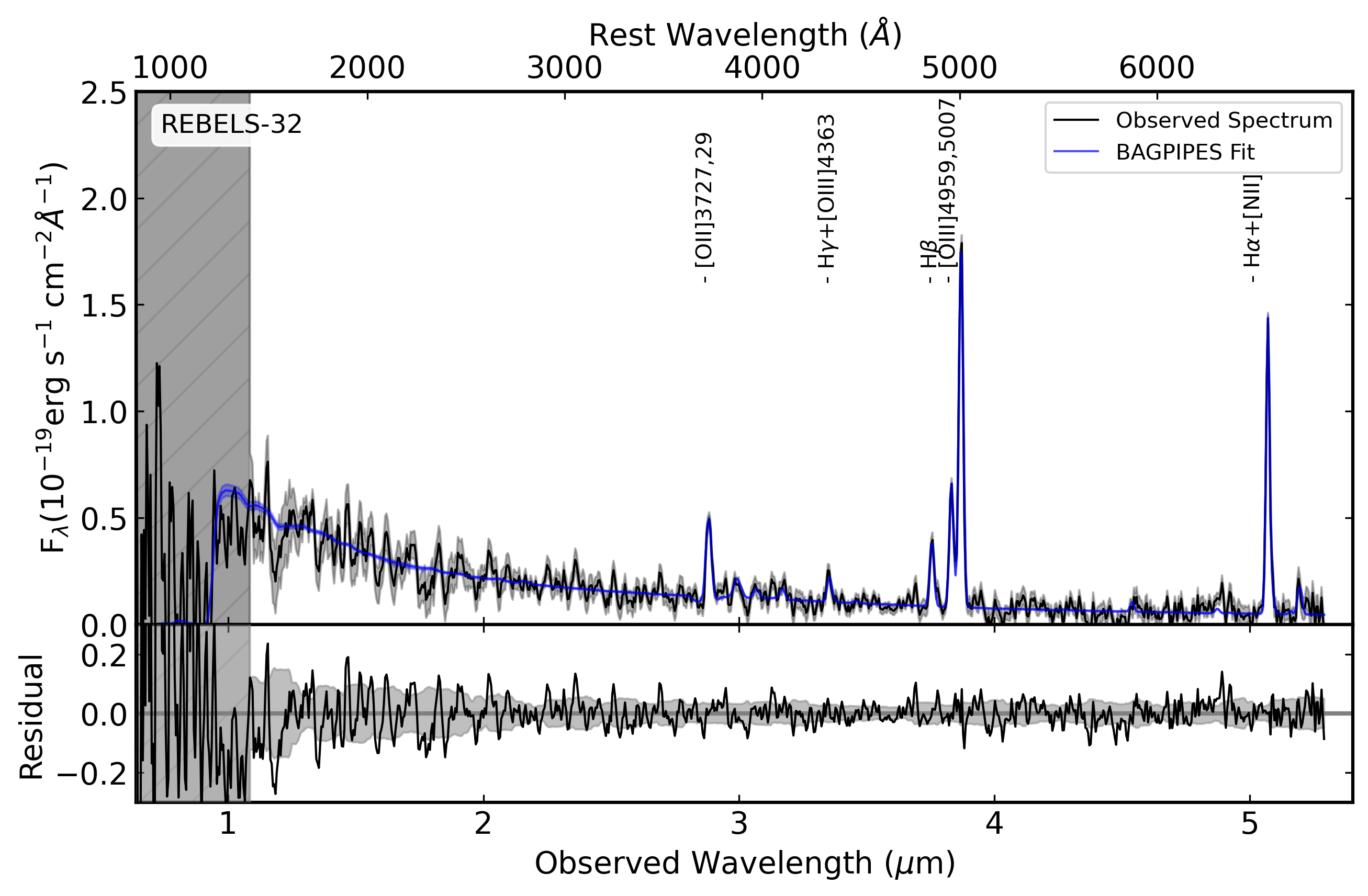} & 
    \includegraphics[width=\columnwidth]{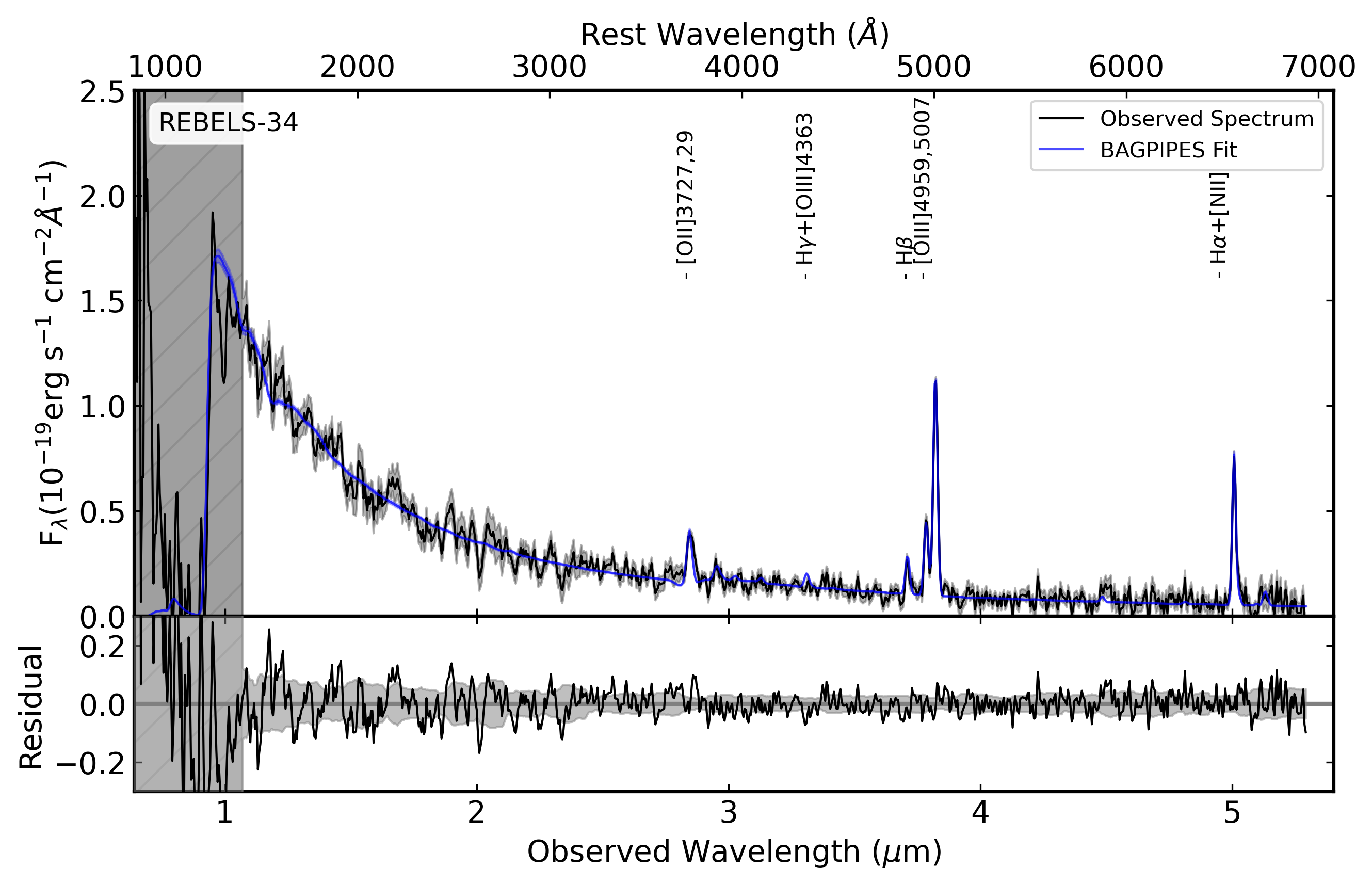} \\ 
    \includegraphics[width=\columnwidth]{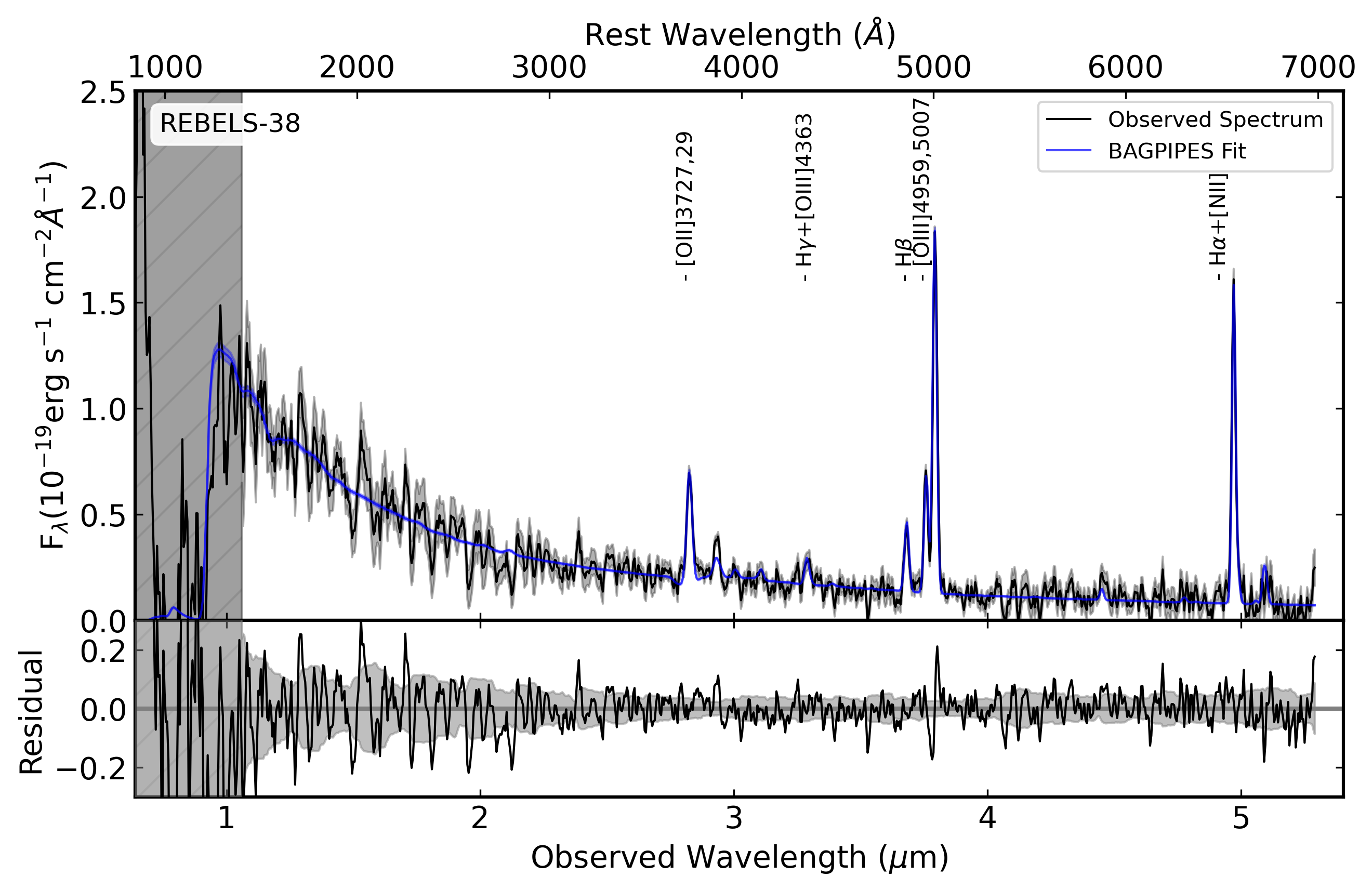} & 
    \includegraphics[width=\columnwidth]{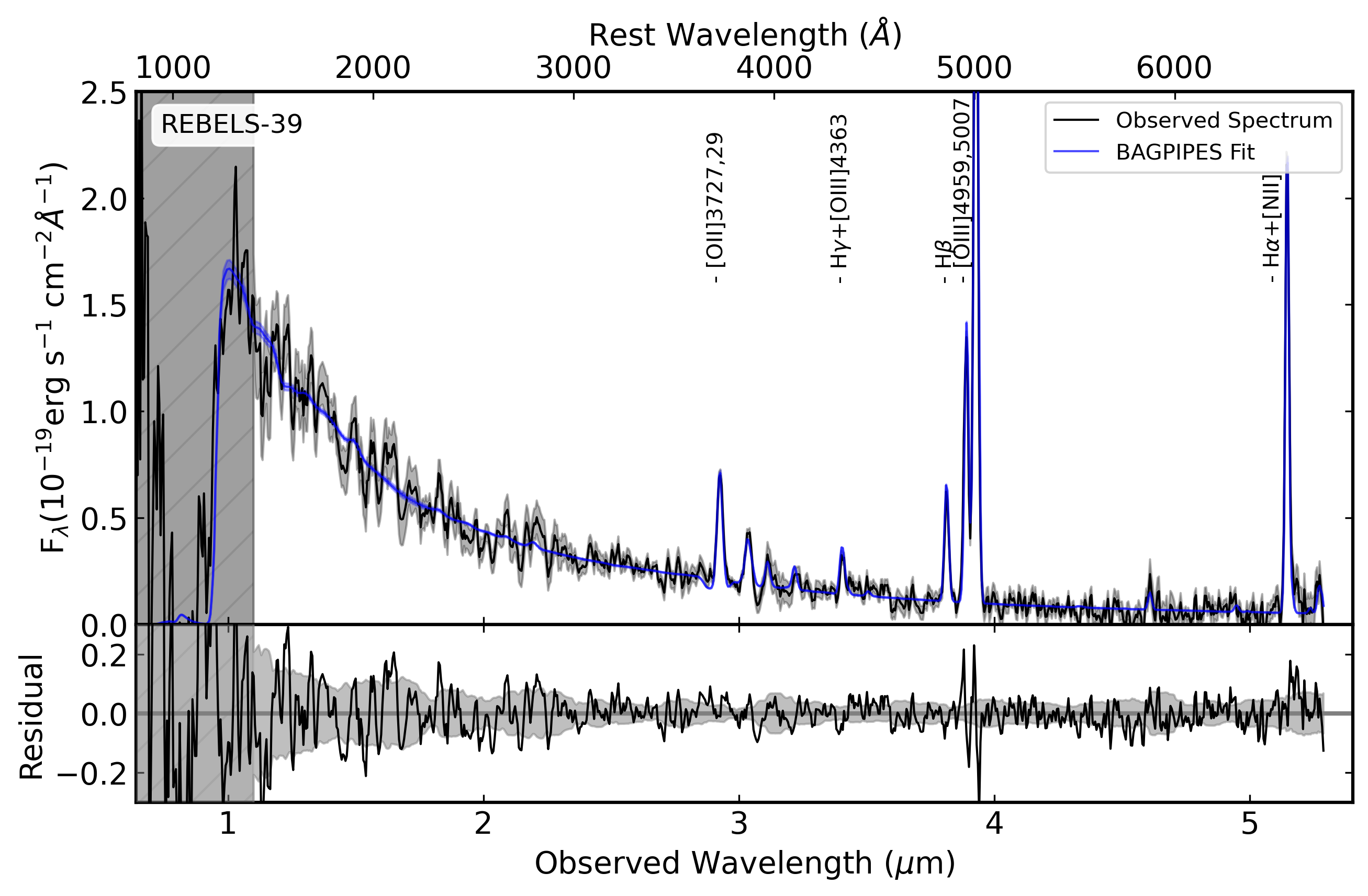} \\ 
    
    \end{tabular}
    \caption{Full NIRSpec PRISM spectra for the REBELS IFU galaxies with the 1$\sigma$ error shown by the grey shaded region.  The best-fitting SED model using a flexible dust attenuation law is shown in blue.  The spectrum is masked below a rest-frame wavelength of 1400~{\AA} to exclude effects caused by potential Ly$\alpha$ damping and instrumental effects at low wavelengths.  The bottom panel in each plot shows the fit residuals.  REBELS-08, REBELS-15, and REBELS-25 show evidence for a 2175~{\AA} dust bump.}
    \label{Spectra_appendix2} 
\end{figure*}

In Fig.~\ref{Spectra_appendix1} and \ref{Spectra_appendix2} we show the full NIRSpec PRISM spectrum for each of the 12 REBELS galaxies in our sample and the best-fitting SED model.  These fits were performed using $\tt{BAGPIPES}$ with a flexible dust attenuation law, as described in Section~\ref{sec:methods}.

\bsp	
\label{lastpage}
\end{document}